# Measurement and assignment of $J$ = 5 to 9 rotational energy levels in the 9070-9370 cm⁻¹ range of methane using optical frequency comb double-resonance spectroscopy

**Adrian Hjältén[1], Vinicius Silva de Oliveira[1], Isak Silander[1], Andrea Rosina[1], Michael Rey[2], Lucile Rutkowski[3], Grzegorz Soboń[4], Kevin K. Lehmann[5], and Aleksandra Foltynowicz[1,*]**

[1] Department of Physics, Umeå University, 901 87 Umeå, Sweden

[2] Groupe de Spectrométrie Moléculaire et Atmosphérique, UMR CNRS 7331, BP 1039, F-51687 Reims Cedex 2, France

[3] Univ Rennes, CNRS, IPR (Institut de Physique de Rennes)-UMR 6251, F-35000 Rennes, France

[4] Faculty of Electronics, Photonics and Microsystems, Wrocław University of Technology, 50-370 Wrocław, Poland

[5] Departments of Chemistry & Physics, University of Virginia, Charlottesville, VA 22904, USA

*aleksandra.foltynowicz@umu.se

**Abstract**: We use optical-optical double-resonance (OODR) spectroscopy with a continuous wave (CW) pump and a cavity-enhanced frequency comb probe to measure high rotational energy levels of methane in the upper part of the triacontad polyad ($P6$). A high-power CW optical parametric oscillator, tunable around 3000 cm⁻¹, is consecutively locked to the P(7, $A_2$), Q(7, $A_2$), R(7, $A_2$), and Q(6, $F_2$) transitions in the $\nu_3$ band, and a comb covering the 5800-6100 cm⁻¹ range probes sub-Doppler ladder-type transitions from the pumped levels with $J'$ = 6 to 8, respectively. We report 118 probe transitions in the $3\nu_3 \leftarrow \nu_3$ spectral range with uncertainties down to 300 kHz (1 × 10⁻⁵ cm⁻¹), reaching 84 unique final states in the 9070-9370 cm⁻¹ range with rotational quantum numbers $J$ between 5 and 9. We assign these states using combination differences and by comparison to theoretical predictions from a new *ab initio*-based effective Hamiltonian and dipole moment operator. This is the first line-by-line experimental verification of theoretical predictions for these hot-band transitions, and we find a better agreement of transition wavenumbers with the new calculations compared to the TheoReTS/HITEMP and ExoMol databases. We also compare the relative intensities and find an overall good agreement with all three sets of predictions. Finally, we report the wavenumbers of 27 transitions in the $2\nu_3$ spectral range, observed as V-type transitions from the ground state, and compare them to the new Hamiltonian, HITRAN2020, ExoMol and the WKMLC line lists.

## 1   Introduction

Accurate models of high-temperature spectra of methane are needed in fields such as astrophysics [1-3] and combustion [4-7]. These models require accurate predictions of highly excited ro-vibrational levels involved in hot-band transitions. Because of the coincidence of the fundamental vibrational mode frequencies and strong near-resonant couplings between them caused by the Fermi, Darling-Dennison, and Coriolis resonances, the energy levels of methane form groups of interacting states called polyads. The $PN$ polyad is the set of states for which $2n_1+n_2+2n_3+n_4$ = N, where $n_i$ is the number of quanta in vibrational normal mode i. The number of sub-levels in each polyad increases exponentially with N [8]. Up to now, the most complete theoretical methane line lists for high-temperature applications have been available in the TheoReTS/HITEMP [9-11] and ExoMol [12, 13] databases. However, little experimental data exists to test their accuracy in the energy range above 7500 cm⁻¹ [14].

The fundamental $\nu_3$ C-H stretch band of methane at 3000 cm⁻¹ and its overtone $2\nu_3$ band at 6000 cm⁻¹ have been well characterized with high resolution using comb-referenced saturation spectroscopy [15-23] and Doppler-limited dual-comb spectroscopy [24, 25]. Extensive room- and low-temperature line lists in the tetradecad ($P4$) region (4760-6250 cm⁻¹) have been reported based on Fourier transform infrared (FTIR) spectroscopy [8, 26-28], cavity ring-down spectroscopy (CRDS) and direct absorption spectroscopy (DAS) [29, 30]. The WKLMC line list based on CRDS and DAS measurements reaches the icosad ($P5$) region (up to 7919 cm$^{-1}$) [31]. At higher wavenumbers, the only available line lists are from low and room temperature FTIR spectroscopy in the upper part of the triacontad ($P6$) and the full tetracontad ($P7$) range (8850-10435 cm$^{-1}$) [32-34]. Most recently, the 10800-14000 cm$^{-1}$ range in the Kitt Peak FTIR spectrum has been analyzed by Campargue *et al.* [35]; this work also includes an extensive summary of available data in this spectral range.

Room- and low-temperature precision spectroscopy of overtone bands provides valuable information about levels that can be reached from the ground vibrational level but, due to selection rules, does not shed light on many of the levels involved in hot-band transitions (i.e., transitions starting from excited vibrational states). High-temperature spectra [36-38], on the other hand, are congested and do not allow line-by-line assignments; only absorption cross-sections can be compared to theory. At room temperature, levels belonging to the $P1$ polyad are thermally populated, allowing observation of resolved hot-band transitions from these states [39]. Resolved hot-band transitions starting from higher energy levels can be observed only under non-equilibrium conditions or using nonlinear spectroscopic techniques. Dudas *et al.* [40] observed transitions in the $P6$ - $P2$ and $P7$ - $P3$ regions under non-local-thermal-equilibrium conditions in a hypersonic flow using CRDS. They assigned 22 hot-band transitions in the 5880-6060 cm$^{-1}$ range, reaching levels in the 8562-9833 cm$^{-1}$ range, mostly within the $2\nu_3 + 2\nu_4$ and $2\nu_3 + 3\nu_4$ bands. These transitions have rotational quantum numbers $J \leq 4$, because the spectra, recorded in hypersonic expansion from a Laval nozzle, are vibrationally excited but remain rotationally cold. The final state assignment was not straightforward and required complex analysis based on combination difference methods and comparison to synthetic models.

A nonlinear technique that allows measurement of transitions from well-defined excited states is optical-optical double-resonance spectroscopy (OODR), in which a high-power laser populates selected ro-vibrational levels, and a probe laser measures ladder-type transitions from each of these levels to highly excited states. The probe transitions are free of Doppler-broadening if the pump has a linewidth narrow enough to interact with only one velocity group of molecules. Using OODR with nanosecond pulsed lasers, de Martino *et al.* measured transitions in the $P6$ - $P4$ ($3\nu_3 \leftarrow 2\nu_3$) range of methane [41-43], with resolution worse than the Doppler broadening of the transitions. Much higher resolution was obtained using OODR spectroscopy with continuous wave (CW) lasers by Okubo *et al.* [44], who used two difference-frequency-generation CW sources to measure ten transitions between Q(1) to Q(4) in the $2\nu_3(A_1) \leftarrow \nu_3$ range with sub-Doppler resolution and kHz uncertainties, reaching states in the 5970-6070 cm$^{-1}$ range of the $P4$ polyad. Around the same time, our group demonstrated OODR spectroscopy using a CW pump and a frequency comb probe, which combines sub-Doppler resolution with much wider spectral coverage (200 cm$^{-1}$) [45]. We used this method to measure multiple hot-band transitions in the $P6$ - $P2$ ($3\nu_3 \leftarrow \nu_3$) range of methane. In the first demonstration, with a methane sample contained in a liquid-nitrogen-cooled single-pass cell, we detected and assigned 36 transitions reaching 32 states in the 8940-9110 cm$^{-1}$ range with rotational quantum numbers $J$ up to 3 [46]. Next, we implemented a room-temperature enhancement cavity to increase the absorption sensitivity for the comb probe [47] and detected 19 new transitions reaching 16 final states in the 8970-9015 cm$^{-1}$ range with rotational quantum numbers $J = 2$ to 4.

All prior measurements of hot-band methane transitions in the $P6$ - $P2$ range [40, 46, 47] addressed rotational states with $J \leq 4$ and found that the final state term values predicted by TheoReTS agreed within their ~1 cm$^{-1}$ uncertainty with the experimental values. However, the strength of vibration-rotation interactions, as well as the size of matrices that must be diagonalized in the calculation of ro-vibrational eigenvalues and the resulting density of states grow with $J$, and it is thus essential to test the predictions for higher $J$ values. This is especially true given that the TheoReTS predictions are used to model spectra of hot methane, where higher $J$ value states dominate the spectra. The current version of TheoReTS [9-11, 48] for methane is based on extensive variational calculations using accurate *ab initio* potential energy (PES) [49] and dipole moment (DMS) [50] surfaces. At room temperature, most of the line positions were corrected up to the *P*4 (tetradecad) polyad using a set of rotation-vibration energy levels computed from an empirical effective Hamiltonian whose parameters were fitted to experimental data [51-53]. The line positions of medium and strong lines in the *P*5 (icosad) polyad were corrected using empirical levels obtained from the assignment of CRDS and DAS spectra recorded at 80 K and 296 K from a variational line list [54]. Only a few lines of the *P*6 (triacontad) polyad were corrected. Unfortunately, for many of the energy levels in the triacontad, the variational calculation may suffer from a lack of convergence, sometimes making clear and unambiguous identification of lines in experimental spectra quite tedious. The same holds for the HITEMP [11] database since it contains TheoReTS.

At the time of our previous OODR measurements [46], the ExoMol line list yielded much worse agreement with the experimental data than TheoReTS. However, the ExoMol methane line list has recently been significantly updated to include 50 billion transitions with wavenumbers up to 12 000 cm$^{-1}$ [13]. The line list was generated through the solution of the nuclear motion Schrödinger equation for an empirically derived PES and a high-level *ab initio* DMS. The PES was constructed by fitting the ro-vibrational energies of $CH_4$ to a set of highly accurate, experimentally derived energies. The Marvel (Measured Active Rotational Vibrational Energy Levels) analysis [14] replaced the predicted ro-vibrational energies with the experimentally derived values for 23 208 states with $J \leq 27$ below 9986 cm$^{-1}$, covering the lowest eight polyads. In the *P*6 - *P*2 range, the only available data was from the two works mentioned above [40, 46], and thus also limited to states with $J \leq 4$.

In this work, we use optical-optical double resonance spectroscopy with a cavity-enhanced frequency comb probe to measure and assign levels in the *P*6 - *P*2 range of methane with rotational quantum numbers ranging from 5 to 9. We use a 3.3 μm continuous wave high-power optical parametric oscillator (OPO) to pump transitions from the $J'' = 7(A_2)$ level in the ground state to ro-vibrational levels with $J' = 6$ to 8 in the $\nu_3$ band, and a frequency comb probe centered at 1.68 μm to probe sub-Doppler ladder-type transitions from these levels, reaching 84 final states in the 9070-9370 cm$^{-1}$ range with rotational quantum numbers $J$ between 5 and 9. We assign the probe transitions using combination differences, i.e., reaching the same state using different combinations of pump and probe frequencies, as well as theoretical predictions from a new effective Hamiltonian [55]. The use of effective Hamiltonians for the modeling of high-resolution spectra is well-established [56]. For isolated polyads with few vibrational bands, very accurate results can be obtained using few parameters, far less than the number of term values that can be accurately predicted. Unlike variational calculations, the small dimensionality of the effective Hamiltonians makes the computation of the energy levels straightforward, even for high $J$ values. For more complex polyads, containing many vibrational bands and numerous degeneracies and quasi-degeneracies, like the *P*6 polyad of methane, missing information on the 'dark' states may lead to a poor determination of resonance coupling parameters. In that case, such models fail to describe properly the main spectral features, with possibly wrong intensity transfers between two or several successive lines. To our knowledge, no empirical effective Hamiltonian was previously developed to model the whole triacontad polyad. Within that context, a novel methodology was recently proposed in Ref. [55] to construct effective Hamiltonians and dipole moment operators from PES and DMS. The results of these new calculations are used to assign the experimental line lists in this work, and compared to assignments using the TheoReTS/HITEMP [11] and ExoMol predictions [13].

Obtaining absolute line intensities from OODR measurements is difficult since the population in the intermediate $\nu_3$ pumped states is not known accurately. However, assuming equal relaxation rates of the upper and lower states of the pump transition, the pumped population in the upper state matches the depletion in the lower state [46]. The latter can be estimated from the area of the so-called V-type OODR transitions that appear in the spectrum in the centers of the Doppler-broadened transitions with the same lower state as the pump transition. Hence, the population of the lower state of the ladder-type transitions is proportional to the area of the observed V-type transitions. This allows the normalization of the ladder-type intensities by the V-type intensities and comparison to theoretical predictions. We detect 22 V-type transitions in our spectra and we compare their positions to 4 line lists, from the effective Hamiltonian, HITRAN2020 [57], ExoMol [13], and WKLMC [29]. We use selected strongest V-type transitions for normalization of the ladder-type transitions.

## 2 Experimental setup and procedures

The principle of the experiment is similar to that of Ref. [47], but the setup has been rebuilt in order to extend the spectral coverage of the probe spectra and simplify long-term averaging. This included the implementation of a narrow-linewidth CW pump with better long-term stability, a longer soliton shifting fiber that allows tuning the comb probe spectrum to lower wavenumbers, a cavity with a broader spectral coverage and lower mirror dispersion, and a comb-cavity locking scheme where both the comb repetition rate, $f_{rep}$, and offset frequency, $f_{ceo}$, are absolutely stabilized.

The experimental setup is shown in Figure 1. The pump is the idler of a high-power (1.5 W) singly-resonant continuous wave optical parametric oscillator (CW-OPO, TOPTICA, TOPO). The pump frequency is stabilized to the center of a Lamb dip in a selected $CH_4$ transition in the $\nu_3$ band using frequency modulation spectroscopy in a reference cell, similar to what was done in Ref. [46]. The OPO seed laser current (narrow-linewidth external cavity diode laser, TOPTICA, CTL PRO) is modulated at $f_{FM}$ = 23 MHz, which results in phase modulation of the idler. A small fraction (20 to 40 mW) of the pump power is sent, using a combination of a half waveplate ($\lambda/2$) and a dichroic mirror (DM), to a 30-cm-long reference cell (Ref. cell) filled with a few tens of mTorr of pure $CH_4$. The cell has a $CaF_2$ input window and a silver mirror at the back. The back-reflected light is picked off by a pellicle beamsplitter (Pell. BS), detected by a fast photodetector (PD), demodulated at $f_{FM}$, and sent to a proportional-integral servo controller. The feedback is sent to the piezoelectric transducer and drive current of the seed laser. The polarization of the pump is linear and its plane of polarization in front of the cavity is controlled using a half-waveplate.

The probe laser is an amplified Er:fiber frequency comb (Menlo Systems, FC1500-250-WG) with $f_{rep}$ = 250 MHz. A polarization-maintaining microstructured silica fiber (PM-MSF) [58], 1-m long, is used to shift the probe spectrum to wavenumbers ranging from 5800 $cm^{-1}$ to 6100 $cm^{-1}$ with up to 250 $cm^{-1}$ of simultaneous bandwidth (evaluated at –10 dB). The center wavenumber of the probe is tuned by adjusting the comb power coupled into the PM-MSF. The light exiting the PM-MSF is coupled to free space via a polarization-maintaining fiber-coupled optical circulator. The polarization of the comb probe at the output of the circulator is slightly elliptical, with 3% of power along the minor axis, which corresponds to an ellipticity of 0.17.

The sample of 200 mTorr of $CH_4$ (Air Liquide, 99.995% purity) is contained in a 60-cm-long cavity (250 MHz free spectral range, $FSR$) resonant for the comb probe, made of two mirrors with 1 m radius of curvature and a YAG substrate (Layertec). The specified mirror reflectivity is ~99.7% in the 5400-6100 $cm^{-1}$ range and the specified group delay dispersion in this range is < 100 fs². The back surface of the mirrors has an antireflection coating in the 2700-3100 $cm^{-1}$ range, resulting in an empty cavity transmission at the pump wavelengths of 94.5%. The pump and probe beams are combined in front of the cavity using a dichroic mirror with the same coatings as the cavity mirrors. The relative pump/probe polarization is adjusted to the so-called magic angle of 54.7° using the last half-wave plate in the pump beam. This eliminates the impact of pump-induced $M_J$ alignment on the probe absorption strength, where $M_J$, is the quantum number for the projection of the total angular momentum on the axis defined by the pump electric field, and allows measuring the intrinsic line strength of the OODR probe transitions [59]. The probe beam is mode-matched to the TEM$_{00}$ mode of the cavity with a Rayleigh range of 45.8 cm and a beam radius at the waist of 0.5 mm. To maximize the spatial overlap between the pump and the probe beams in the cavity, the pump beam is set to have the waist in the middle of the cavity and the same Rayleigh range as the probe beam. This results in a pump beam radius of 0.7 mm at the waist. The pump passes once through the cavity, and the transmitted power is monitored using a power meter after another dichroic mirror that separates the pump and probe beams after the cavity.

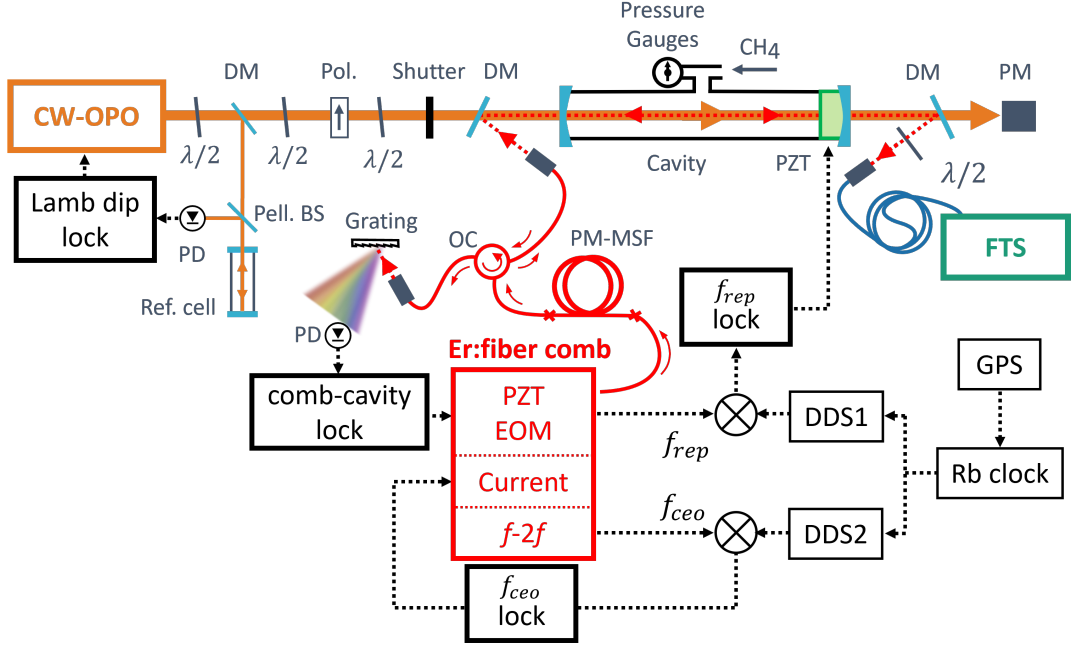

Figure 1. Experimental setup. CW-OPO: continuous wave optical parametric oscillator. $\lambda/2$: half-wave plates. DM: dichroic mirrors. Pol: polarizer. PD: photodiodes. Pell. BS: pellicle beam splitter. Ref. cell: reference cell. PZT: piezoelectric transducer. PM: power meter. FTS: Fourier transform spectrometer. PM-MSF: polarization-maintaining microstructured silica fiber. OC: optical circulator. Current: current input of the comb oscillator. EOM: intracavity electro-optic modulator. $f$-$2f$: an $f$-$2f$ interferometer. DDS: direct digital synthesizers. See text for details.

In our previous work [47], the comb was locked to the cavity using the two-point Pound-Drever-Hall (PDH) method [60, 61]. The $f_{rep}$ was absolutely stabilized to an RF reference, while the $f_{ceo}$ was only indirectly locked via the comb-cavity lock and drifted slightly, which complicated long-term averaging, since the frequency grid was different in consecutive measurements. Here, we modified the locking scheme so that both $f_{rep}$ and $f_{ceo}$ are absolutely stabilized during the measurement, to make the long-term averaging more straightforward. The comb is now PDH locked to the cavity at only one spectral point via feedback to the comb $f_{rep}$, and the $f_{ceo}$ is RF-stabilized to a value that optimizes the cavity transmission. To obtain the PDH error signal, the comb is phase-modulated at 20 MHz using a fiberized electro-optic modulator (EOM) inserted between the oscillator and the amplifier. The light reflected from the cavity is collected by the fiber optical circulator and directed onto a free-space reflection grating that spatially disperses the comb spectrum. A selected part of the spectrum is incident on a high-bandwidth InGaAs photodiode and demodulated at the modulation frequency to yield a PDH error signal. The feedback is sent to an intracavity piezoelectric (PZT) actuator and EOM inside the comb oscillator cavity that acts on the $f_{rep}$. The comb modes around the locking point are transmitted through the cavity without any phase offset. To absolutely stabilize the $f_{rep}$, an error signal is generated by comparing $f_{rep}$ to an RF signal from a tunable direct digital synthesizer (DDS1), which is referenced to a 10 MHz signal from a GPS-disciplined rubidium clock. The feedback is sent to a PZT actuator that controls the sample cavity length by displacing one of its mirrors. Finally, the $f_{ceo}$ is tuned to maximize the bandwidth transmitted through the cavity. The $f_{ceo}$ beatnote is detected using an $f$-$2f$ interferometer and stabilized at the optimum value by locking it to a second Rb-referenced tunable direct digital synthesizer (DDS2) via feeding back to the pump diode current of the comb oscillator.

The comb light transmitted through the cavity is coupled through a polarization-maintaining fiber patch cable to a fast-scanning Fourier transform spectrometer (FTS) with an auto-balancing InGaAs detector [62]. The optical path difference in the FTS is measured using a frequency-stabilized 633 nm HeNe laser (Sios, SL/02/1, fractional frequency stability of $5 \times 10^{-9}$ over 1 h). We use the sub-nominal sampling-interleaving method [63, 64] to acquire spectra with comb-mode limited resolution. We record spectra at 130 $f_{rep}$ values differing by 2.75 Hz by tuning the frequency of the reference signal from DDS1. This corresponds to ~2 MHz steps of the comb modes in the optical domain. At each $f_{rep}$ value we record two interferograms using a digital oscilloscope at 5 Msample/s and 20-bit resolution, one with the pump excitation and one without. To do that, the pump beam is blocked and unblocked on consecutive FTS scans using a shutter. The nominal resolution of the FTS is matched to the $f_{rep}$ and one interferogram is recorded in 2.75 s. The total acquisition time of a full scan of 130 $f_{rep}$ values is 13 min, including dead time (8.3%) for resampling the comb interferogram at the zero crossings and the maxima of the CW reference laser interferogram and saving the data. For averaging, multiple acquisition series are made while scanning the $f_{rep}$ in alternating directions.

We measured spectra with the pump locked to 3 transitions in the $\nu_3$ fundamental band of methane with lower state rotational quantum number $J'' = 7$ and $A_2$ rovibrational symmetry, i.e., P(7, $A_2$), Q(7, $A_2$) and R(7, $A_2$) transitions. The $\nu_3$ Q(7, $A_2$) transition is separated from the neighboring $\nu_3$ Q(6, $F_2$) transition by 256 MHz, which is less than their Doppler width, so the pump locked to the $\nu_3$ Q(7, $A_2$) transition interacts also with a small fraction of molecules in the $J'' = (6, F_2)$ state. Therefore, we also measured a spectrum with pumped locked to the $\nu_3$ Q(6, $F_2$) transition, which we then used in our analysis to look for $J'' = 6$ lines in the $J'' = (7, A_2)$-pumped spectrum. The other pump transitions, P(7, $A_2$) and R(7, $A_2$), are separated from their nearest neighbor by much more than the Doppler width.

Table I lists the center wavenumbers and assignments of the four pump transitions (from Refs [17, 18]), as well as other parameters of the measurement series. The center frequency of the probe spectrum was shifted between the measurements to compensate for the frequency difference between the pump transitions, in order to reach final states with overlapping term value ranges. The probe spectrum incident on the cavity was adjusted by changing the power coupled into the PM-MSF. We note that a lower limit of 5800 cm$^{-1}$ in the spectral coverage of the probe is imposed by the cut-off of the responsivity of the auto-balanced InGaAs photodetector in the FTS. The transmission through the cavity was optimized by adjusting the PDH locking point and the RF frequency of the $f_{ceo}$ beat note. The PDH locking point was set close to the maximum of the incident spectrum by tuning the grating in front of the PDH detector. After the PDH feedback is turned on, the $f_{ceo}$ is manually adjusted to yield the largest transmitted bandwidth through the cavity and locked to the optimum value using DDS2. The feedback to the sample cavity PZT is turned on at the last step to complete the frequency stabilization by absolutely locking $f_{rep}$. During the measurements, the $f_{rep}$ and $f_{ceo}$ are monitored using a counter with 1 s gate time. The standard deviation of the $f_{ceo}$ was 100 mHz for all datasets during the measurement time. We note that the frequency offset between the comb lines and the cavity modes is zero at the PDH locking point, and deviates from zero away from the locking point since the *FSR* of the cavity varies with optical frequency because of the dispersion in the cavity mirror coatings and sample gas. The non-zero comb-cavity frequency (phase) offset combined with the sample resonance dispersion causes an asymmetry in the absorption line profiles in cavity transmission. This effect is well understood and included in the cavity transmission function [61] and does not affect the accuracy of the determination of the line positions.

Table I. Parameters of the four measurement series. Column 1: Pump transition assignment including lower and upper state rotational labels and counting numbers from HITRAN, and wavenumbers taken from Refs [17, 18]. Column 2: The transmitted probe bandwidth evaluated at -10 dB. Column 3: Wavenumber of the PDH locking point, i.e., the point in the spectrum where the comb-cavity offset is zero. Column 4: Sample pressure in the cavity. Column 5: Sample pressure in the reference cell used for the pump frequency stabilization. Column 6: Pump power incident on the sample (calculated after the first cavity mirror). Column 7: On-resonance pump transmission, i.e. pump power transmitted through the cavity when it is on resonance with the pump transition divided by the pump power when off resonance. Column 8: Number of $f_{rep}$ scans recorded for averaging.

| Pump transition | Comb probe parameters | | Sample pressure | | Pump parameters | | |
|---|---|---|---|---|---|---|---|
| $v_3$ band: assignment and wavenumber [cm$^{-1}$] | Coverage [cm$^{-1}$] | PDH locking point [cm$^{-1}$] | Cavity [mTorr] | Reference cell [mTorr] | Incident power [mW] | On-resonance transmission | # averages |
| P(7, $A_2$) $7A_2$ (1) → $6A_1$ (10) 2948.10794477(8) | 5910 – 6095 | 6035 | 200 | 64 | 910 | 8% | 8 |
| Q(7, $A_2$) $7A_2$ (1) → $7A_1$ (9) 3016.49766637(6) | 5865 – 6080 | 5980 | 200 | 52 | 960 | 6% | 8 |
| R(7, $A_2$) $7A_2$ (1) → $8A_1$ (11) 3095.17923673(13) | 5820 – 5980 | 5932 | 200 | 43 | 900 | 3% | 10 |
| Q(6, $F_2$) $6F_2$ (1) → $6F_1$ (22) 3016.48912913(11) | 5840 – 5980 | 5900 | 200 | 58 | 940 | 9% | 10 |

## 3 Data analysis

The vibrational energy levels and transitions addressed by the CW pump and comb probe are shown schematically in Figure 2a) for the case of pumping from the $J'' = (7, A_2)$ state. When the pump beam is blocked, the probe interacts only with the Doppler-broadened transitions in the $2v_3$ overtone region, which also contains a number of combination bands. When the pump beam is on, two types of sub-Doppler OODR transitions appear: ladder-type transitions to the levels in the $3v_3$ overtone region, as well as V-type transitions in the centers of the Doppler-broadened transitions that share the lower level with the pump transition. We note that because of the cavity enhancement, the strongest Doppler-broadened transitions absorb all the comb light. We also observe a few percent decrease in the absorption of the Doppler-broadened transitions when the pump is on, which we attribute to the change of sample density due to heating by the pump. Moreover, the intensities of some Doppler-broadened lines increase, and new Doppler-broadened lines appear in the spectrum as a result of collisional redistribution of the population transferred by the pump, which increases the population in states that are subsequently probed by the comb. These transitions are known as 4-level OODR transitions (i.e., four distinct energy levels are involved in the pump and probe transitions), as distinct from the 3-level OODR transitions, where the pump and probe transition share a common state. These additional 4-level OODR features lack the sub-Doppler component and thus are easily distinguished from the 3-level OODR transitions that are the focus of the present work.

Figure 2b) shows a more detailed schematic of the ladder-type OODR transitions addressed by the different pump and probe combinations starting from the $J'' = (7, A_2)$ state, including the rotational sub-levels. The pump was locked to the P(7, $A_2$), Q(7, $A_2$) and R(7, $A_2$) transitions of the $v_3$ band, which results in the lower levels of the ladder-type transitions having rotational quantum numbers $J'$ of 6, 7, and 8, respectively. Hence, the rotational quantum numbers $J$ of the final states reached by the ladder-type OODR transitions are between 5 and 9 (all of $A_2$ symmetry). Notably, states with $J = 6 - 8$ can be reached by more than one pump-probe combination, enabling a firm assignment of the final state $J$ number by the so-called combination difference method. The final state term values are the sum of the term value for the ground vibrational state $J = (7, A_2)$ level, and the wavenumbers of the pump and probe transitions.

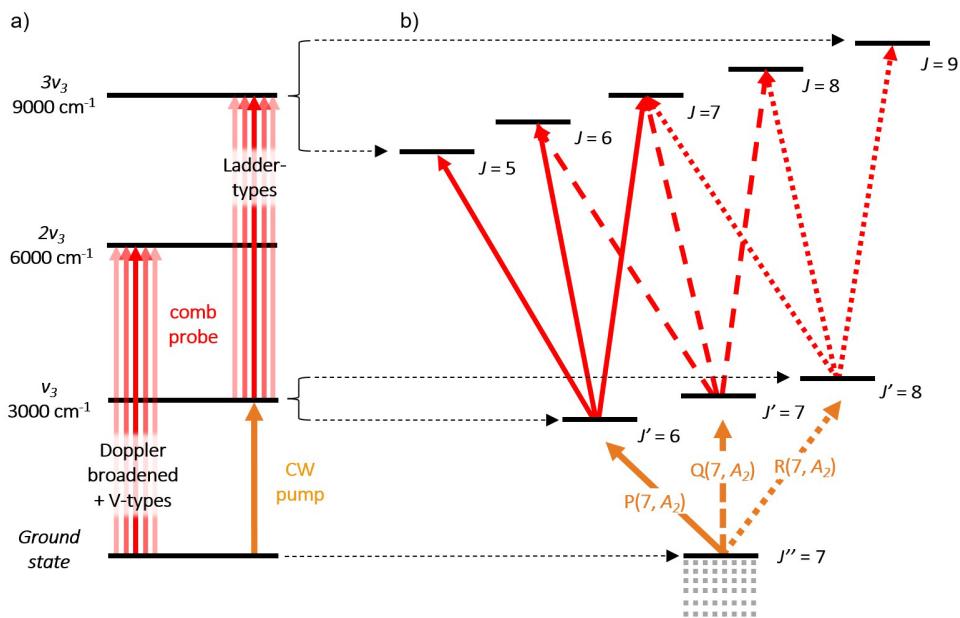

Figure 2. a) Simplified depiction of the vibrational levels addressed by the CW pump (orange) and comb probe (red). The pump excites a narrow velocity group of molecules to a selected rotational sub-level of the $v_3$ band. The probe then excites hot-band transitions from the pumped level. The probe also excites Doppler-broadened transitions from the ground state to the $2v_3$ region, even in the absence of the pump. In the presence of the pump, V-type transitions appear in the centers of the Doppler transitions sharing the lower level with the pump. b) A more detailed schematic of the rotational-vibrational states addressed by the pump and probe lasers starting from the $J'' = (7, A_2)$ state. The pump populates a selected intermediate state of the $v_3$ vibration with $J' = 6$, 7 or 8 for the pump transitions P(7, $A_2$) (orange solid line), Q(7, $A_2$) (orange dashed line) and R(7, $A_2$) (orange dotted line), respectively. The probe subsequently induces transitions to final states with $J$ between 5 and 7 for P(7, $A_2$) pump (red solid lines), between 6 and 8 for Q(7, $A_2$) pump (red dashed lines), and between 7 and 9 for R(7, $A_2$) pump (red dotted lines).

Sections 3.1 and 3.2 describe the different stages of data analysis. First, to determine the cavity finesse and quantify the change of absorption caused by heating of the sample by the pump, we analyzed the Doppler-broadened lines in the background spectra (pump blocked) and OODR spectra (pump unblocked). For this, we baseline-corrected and interleaved the spectra to ~20 MHz sampling point spacing, sufficient to resolve the Doppler-broadened transitions. Next, to detect and analyze the sub-Doppler OODR transitions, we normalized the OODR spectra to the background spectra, and interleaved them to ~2 MHz sampling point spacing. Finally, for quantification of the intensities of the ladder-type transitions, we analyzed selected sub-Doppler V-type transitions in the OODR spectra interleaved to the ~2 MHz sampling point spacing.

### 3.1 Cavity finesse and sample density

We retrieved the cavity finesse as a function of wavenumber from fits to the Doppler-broadened lines in the background spectra. Prior to fitting, we removed the baseline originating from the comb spectral envelope using the procedure described in Appendix A, and then averaged and interleaved the spectra to 20 MHz sampling point spacing. We divided these interleaved, baseline-corrected and averaged spectra into 3-cm$^{-1}$-wide segments and fit the cavity transmission function [61] to each segment. In the fit, the Doppler-broadened lines were modeled using a Gaussian line shape function with line parameters fixed to the HITRAN2020 [57] values, methane fraction fixed to 1, and with the cavity finesse and the comb-cavity phase offset as free parameters, but constrained to constant over each segment. We also fitted baseline components with periods larger than 0.6 cm$^{-1}$ using the cepstral method [65]. We excluded the segments in which the Doppler-broadened absorption was too strong (transmission <10% for the strongest lines) or too weak (signal-to-noise ratio, SNR, for the strongest lines <15). The finesse values retrieved from the Q(7, $A_2$)-pumped spectrum were systematically lower than the finesse values retrieved from the P(7, $A_2$)- and R(7, $A_2$)-pumped spectra. This was most likely due to the methane sample being diluted by impurities in the gas supply line during the Q(7, $A_2$)-pumped measurement. Therefore, we first fitted a third order polynomial to the finesse values retrieved from the P(7, $A_2$)- and R(7, $A_2$)-pumped spectra, and then adjusted a constant correction factor for the finesse values retrieved from the Q(7, $A_2$)-pumped spectrum to minimize the standard deviation of the residuals between these values and the fit to the values from the P(7, $A_2$)- and R(7, $A_2$)-pumped data. This correction factor was found to be 1.1, indicating a 10% decrease in sample concentration. Finally, we combined the three finesse data sets (with Q(7, $A_2$)-pumped results multiplied by 1.1), and repeated the $3^{rd}$ order polynomial fit, which is shown in Figure 3. The relative uncertainty in the finesse is 4%, determined from the 1σ confidence interval of the fit, indicated by the shaded region. Similarly, we performed a polynomial fit to the comb-cavity offset values retrieved from each spectrum.

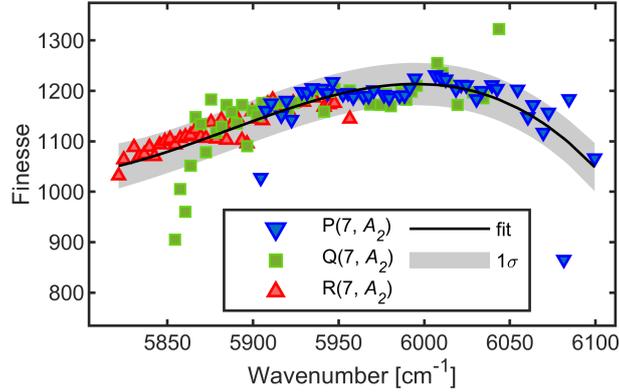

Figure 3. The cavity finesse retrieved from fits to the Doppler-broadened lines in the three interleaved and baseline-corrected background spectra, with pump on the P(7, $A_2$), Q(7, $A_2$) and R(7, $A_2$) transition (downward blue triangles, green squares, upward red triangles, respectively), together with a polynomial fit (black curve) and 1σ confidence interval (shaded region). The error bars on the individual finesse values are negligible on this scale.

To quantify the effect of heating by the pump, we corrected the baseline, interleaved and averaged the OODR spectra (with the pump unblocked) in a similar way as the background spectra. We then performed a global fit to all Doppler-broadened lines in the background and OODR spectra, with the finesse and the comb-cavity offset fixed to the values from the polynomial fits described above, and with one scaling factor for the absorption intensity of the Doppler-broadened lines as a free parameter. Before the fit, we masked regions of ±1 cm⁻¹ around Doppler-broadened lines with transmission lower than 20%, regions of ±0.01 cm⁻¹ around the ladder-type transitions, as well as the edges of the spectra where the standard deviation of the noise was larger than 0.01. We found that the absorption in the three $J = (7, A_2)$-pumped spectra was reduced by ~5% compared to when the pump was blocked. This absorption decrease matches what is expected from the sample density decrease caused by heating by the pump beam, which we estimate to about 15 K based on the geometry of the cavity and the pump beam, and the absorbed pump power. In order to compensate for the heating, we corrected all OODR line intensities by multiplying them with factors of 1.05, 1.20 and 1.06 for the P(7, $A_2$)-, Q(7, $A_2$)- and R(7, $A_2$)-pumped measurements, respectively, where the correction factor for the Q(7, $A_2$)-pumped measurement also compensated for the sample dilution.

## 3.2 Sub-Doppler OODR transitions

To detect and analyze the OODR transitions, we normalized the OODR spectra to the background spectra at each $f_{rep}$ step, which largely cancels the baseline originating from the comb envelope, as well as removes most of the Doppler-broadened absorption features. Ideally, the normalized transmission spectrum would contain only the sub-Doppler OODR lines. Here, however, the cancellation of the Doppler-broadened lines in the $2\nu_3$ region is not complete since heating by the pump reduces the overall absorption in the OODR spectra compared to the background spectra (see Section 3.1). This means that a few percent of each Doppler-broadened line is left in the normalized spectrum, which is above the noise level for the strongest unsaturated lines. The normalized spectrum also contains the Doppler-broadened lines whose intensities increase because of the redistribution of population due to thermal collisions, as well as the Doppler-broadened 4-level OODR transitions. Moreover, at the positions of the strongest Doppler-broadened lines that absorb almost all light, the normalized OODR spectrum is noisy. Finally, because of the scan-to-scan fluctuations of the transmitted comb envelope, a slowly varying baseline is present in the normalized spectra. We removed this baseline and the remaining Doppler-broadened lines as described in Appendix B, averaged and interleaved the spectra to a point spacing of 2 MHz. To localize the sub-Doppler OODR transitions, we ran a peak detection routine described in Section 3.2.1 below, and then retrieved the parameters of all these transitions by fitting a model described in Sections 3.2.2 and 3.2.3.

### 3.2.1 Line detection

We searched for the sub-Doppler ladder-type and V-type OODR transitions in the interleaved baseline-corrected normalized spectra using a custom peak detection routine. First, we convolved the interleaved normalized spectrum with a Lorentzian dispersion line shape with half-width at half-maximum (HWHM) of 10 MHz and took a derivative of this convolution by subtracting the neighboring points. Next, we calculated a 50-point moving average of the original spectrum and subtracted it from the original spectrum to remove features with an HWHM larger than 100 MHz. Finally, we multiplied the two processed spectra, yielding a spectrum with positive peaks at the positions of both ladder-type and V-type transitions. We searched for these peaks using the MATLAB *findpeaks* function. To reject false detections from the noisy areas (e.g., where the Doppler-broadened lines absorb all light), we used the MATLAB *islocalmax* function to find the number of local maxima within a 1 GHz range around a given peak. If the peak was well isolated, it was the only local maximum (or one of a few) in the range. By tuning the prominence limit of this function, the sensitivity of the peak finder was tuned until no new lines were detected and approx. 50% of the detected peaks were false. We inspected all peaks by eye and rejected these false detections, e.g., when the peaks were very noisy or did not have a clear shape. Finally, we sorted the peaks into ladder- and V-type based on their sign in the original spectrum. Some ladder-type detections were removed later during the fitting process if they had an SNR below 5.

The spectra recorded with the pump locked to the $\nu_3 Q(6, F_2)$ and $\nu_3 Q(7, A_2)$ transitions contain ladder-type transitions from two $\nu_3$ levels with $J' = (6, F_2)$ and $(7, A_2)$, because these pump transitions overlap within their Doppler width. However, in the $Q(7, A_2)$-pumped spectra, the ladder-type transitions from the $J' = (6, F_2)$ level appear as two weaker peaks separated by ~1 GHz, i.e., ~4 times the frequency difference of the two pump transitions, where one factor of 2 comes from the difference in absorption frequency of the two opposite velocity groups interacting with the pump in the $Q(6, F_2)$ transition, and the second factor of ~2 comes from the ratio of the probe to pump frequencies. The same is true for the transitions starting from $J' = (7, A_2)$ appearing in the $Q(6, F_2)$-pumped spectrum. Thus, we removed doublets of lines with equal intensity separated by ~1 GHz from the list of peaks in the Q-pumped spectra, especially if a corresponding single line centered between the doublet appeared in the other spectrum.

### 3.2.2 Ladder-type line fitting

We retrieved the parameters of the sub-Doppler ladder-type transitions by fitting the cavity transmission function [61] to the averaged interleaved baseline-corrected normalized spectra in windows of ±1 GHz centered around each ladder-type line using the MATLAB *fit* function. Each line was modeled as a sum of a narrow Lorentzian line shape for the sub-Doppler component, and a broader Gaussian line shape with the same center frequency to account for the redistribution of the pumped velocity group through elastic collisions. The cavity finesse was fixed to the values obtained from the polynomial fit described in Section 3.1. The cavity transmission function also included the Doppler-broadened transitions from the ground state simulated using the HITRAN2020 [57] line parameters. We then divided this function by the transmission function containing only the Doppler-broadened lines, to account for the normalization step. This is necessary because even though the Doppler-broadened lines are removed by normalization, their presence in the spectrum modifies the effective cavity enhancement of the interaction length with the sample, as well as the intracavity dispersion. In the fit, the free parameters were the center frequency, common for the Lorentzian and Gaussian components, the Lorentzian and Gaussian widths, the integrated absorption coefficients, and the comb-cavity phase offset. Fitting the comb-cavity offset, instead of fixing it to the values obtained from fits to the Doppler-broadened lines described in Section 3.1, yielded better precision on the fitted line shapes. To account for the remaining slight baseline distortions, a polynomial baseline was fitted together with the model. In most cases, a 1st order polynomial was sufficient, but sometimes a 3rd order polynomial was required. For some lines, we reduced the fit window size to avoid baseline problems on the wings of the lines, due to features not cancelled by the normalization or removed by fitting. 6-MHz-wide windows around the sidebands at ±46 MHz originating from the phase modulation of the pump were masked during the fitting process (the factor of two compared to the modulation frequency of $f_{FM} = 23$ MHz comes from the ratio of probe over pump frequencies). Figure 4 shows an example of a ladder-type line observed at 5956.54279(1) cm$^{-1}$ in the P(7, $A_2$)-pumped spectrum (black solid curve), together with the fit of the full model shown as the red solid curve and the Lorentzian and Gaussian components shown separately as dashed blue and green curves, respectively. The lower panel shows the model residual. For lines where the Gaussian component had an SNR below 5, we fixed the Gaussian-to-Lorentzian intensity ratio to 1.6 and the Gaussian HWHM to 216 MHz, which were the mean values found from fits to seven strongest lines with SNR > 10 for the Gaussian component and flat residuals. This prevented the low-SNR fits from yielding unphysical values for the Gaussian parameters.

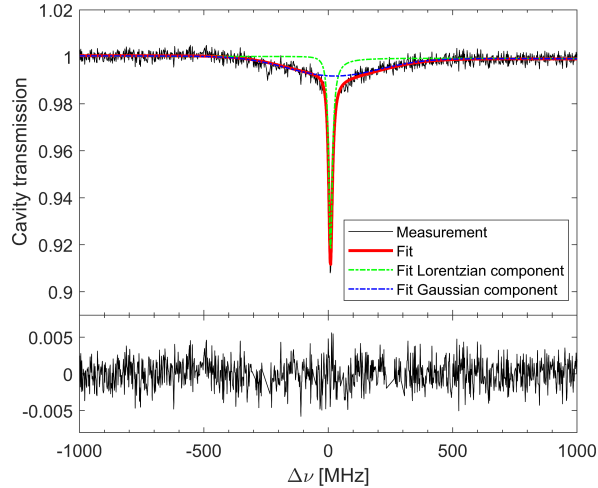

Figure 4. The sub-Doppler ladder-type OODR transition at 5956.54279(1) cm$^{-1}$ in the P(7,$A_2$)-pumped spectrum (black) and the fit (red solid curve) plotted as a function of the detuning, $\Delta\nu$, from the center of the transition. The dashed curves show the Gaussian (blue) and Lorentzian (green) components of the fit respectively, and the bottom panel shows the residuals.

### 3.2.3 *V-type line fitting*

To find the center frequencies of the V-type transitions, we fitted them in the averaged interleaved baseline-corrected normalized spectra using the model described in Section 3.2.2, with inverted sign, since the V-type peaks point in the opposite direction. Because all V-type transitions are relatively weak, we fixed the Gaussian width and the ratio between the Lorentzian and Gaussian integrated absorption coefficients to the mean values obtained from fits to the strongest ladder-type lines (see

Section 3.2.2). In addition, we also fixed the Lorentzian width to the mean value of the HWHM of the ladder-type transitions, i.e., 8 MHz.

For quantification of ladder-type intensities (see Section 4.5), we selected one V-type transition in each $J'' = (7, A_2)$ - pumped spectrum with the highest SNR and the same change of rotational quantum number as the pump transition. To accurately retrieve the intensities of these V-type transitions, we analyzed them in the baseline-corrected and averaged OODR spectra (see Section 3.1) interleaved to 2 MHz sampling point spacing. This allows seeing the V-type transition on top of the corresponding Doppler-broadened transition and with a $2^{1/2}$ times higher SNR than in the normalized spectrum. Figure 5a) shows the Q(7, $A_2$) transition of the $2\nu_3$ band displaying a V-type feature in the $\nu_3$ Q(7, $A_2$)-pumped spectrum. The V-type feature appears not to be in the center of the Doppler-broadened line because of the asymmetry caused by a non-zero comb-cavity phase offset at this wavenumber. Again, this asymmetry is included in the model and does not affect the determination of the center frequency.

To find the intensities of the three selected V-type transitions, we fit the cavity transmission function with a model consisting of a Gaussian function for the Doppler-broadened line (with Doppler width fixed to the value calculated for the respective line position, ~275 MHz HWHM), and a sum of a Lorentzian and a Gaussian component for the V-type feature, similar to what was done for the ladder-type transitions (with the opposite sign). We again fixed the width of the Gaussian component of the V-type transition, the intensity ratios of the two components, and the Lorentzian width to the mean values obtained from fits to the ladder-type transitions. The fitting parameters were the center frequency, common for all line shape components, the integrated absorption of both the V-type and the Doppler-broadened line, as well as the comb-cavity phase offset. The fits were applied in windows of ±100 MHz around the V-types transitions, as shown in Figure 5b).

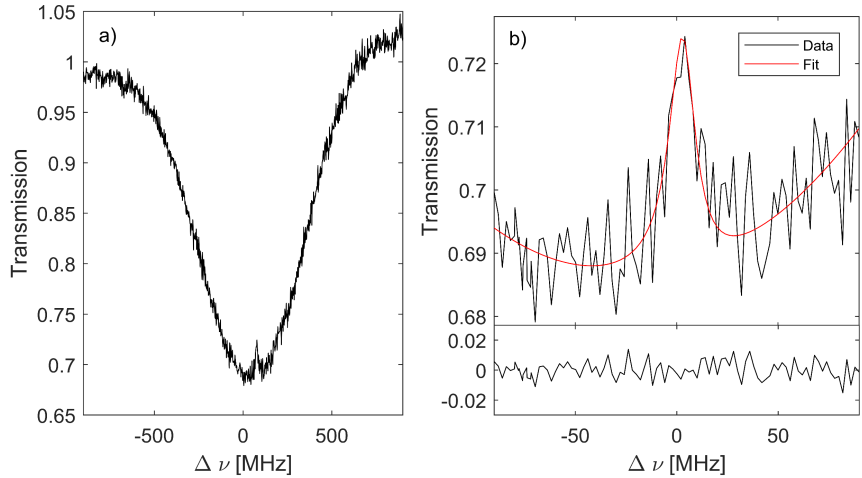

Figure 5. a) The $2\nu_3$ Q(7,$A_2$) line with the V-type transition in the Q(7, $A_2$)-pumped spectrum. b) A zoom-in on the V-type transition (black) and the corresponding fit (red). Residuals are shown in the bottom panel.

### 3.2.4   Frequency uncertainty

The sub-nominal resolution procedure [63, 64] relies on matching the FTS spectral sampling points to the comb mode positions. This is done in post-processing by adjusting the value of the wavelength of the CW reference laser, $\lambda_{ref}$, used for calibration of the optical path difference to minimize the instrumental line shape in the spectra [64]. In principle, the center frequencies of the sub-Doppler lines should be independent of the value of $\lambda_{ref}$ (once $\lambda_{ref}$ is known with precision better than $f_{rep}$), since their width is narrower than $f_{rep}$ [64]. Here, however, each sub-Doppler line resides on top of a Doppler-broadened collision-induced component whose width is comparable to the $f_{rep}$ of the comb. This causes an apparent shift of the center frequency of the OODR lines with $\lambda_{ref}$. To estimate this effect, we analyzed the spectra of the strongest ladder-type OODR transition in each spectrum assuming different $\lambda_{ref}$ values. We fit the model described in Section 3.2.2 to these spectra and found the optimum $\lambda_{ref}$ as the one that yields the minimum of the fit residuals, corresponding to the smallest instrumental line shape function. We then checked how the fitted center frequency changes with $\lambda_{ref}$, similarly to what was done in Refs [39, 47]. From the slope of this dependence and the uncertainty on the optimum $\lambda_{ref}$, we estimated the contribution of the sub-nominal resolution procedure to the uncertainty in the line positions to be 330 kHz.

We estimated the influence of the pressure and Stark shift on the probe transition frequencies from literature values for other $CH_4$ ro-vibrational transitions. Lyulin et al. [66] reported a self-induced pressure shift coefficient of $(-0.017 \pm 0.003)$ cm$^{-1}$/atm$^{-1}$ for methane lines in the 6000 cm$^{-1}$ range, which yields a $-130$ kHz pressure shift for the probe lines at 200 mTorr, which we included in the uncertainty budget. Okubo et al. [17] reported a $(-13 \pm 17)$ kHz/W power shift coefficient for a sub-Doppler Lamb dip in the P(7,$E$) line of the $\nu_3$ band and a beam radius of 0.71 mm, similar to the beam radius of our pump. For the pump power used in our experiment, i.e. 900 mW, this results in a $(-11 \pm 15)$ kHz shift, which is negligible compared to the pressure shift and the uncertainty from the $\lambda_{ref}$ calibration.

### 3.3  Absorption sensitivity

To estimate the absorption sensitivity, we calculated the noise as the standard deviation of the residuals of the ladder-type OODR line fits, multiplied by the square root of the number of measurements to allow comparison between datasets with different number of averages. We excluded from the calculations the lines at the edges of the spectra (approx. 20 cm$^{-1}$ at each edge of the spectra) where the noise increases by more than twice the mean value. Windows of $\pm50$ MHz around the center of each line were also excluded from the calculation as they reflect the mismatch between the model and the data and the phase modulation sidebands, and not the measurement noise. The mean value of the noise, $\sigma$, was found to be 0.0081, 0.0085, and 0.0176 for the P(7, $A_2$)-, Q(7, $A_2$)- and R(7, $A_2$)-pumped spectra, respectively. The noise was higher in the R(7, $A_2$)-pumped spectrum because of the lower responsivity of the InGaAs photodiodes in the FTS detector at the lower wavenumbers. The minimum detectable absorption coefficient, $\alpha_{min} = \sigma/L_{eff}$, in one normalized interleaved spectrum acquired in 13 minutes is thus $1.8 \times 10^{-7}$ cm$^{-1}$ for the P(7, $A_2$)- and Q(7, $A_2$)-pumped spectra, and $4 \times 10^{-7}$ cm$^{-1}$ for the R(7, $A_2$)-pumped spectrum. Here, $L_{eff} = 2FL/\pi$ is the effective cavity length, with the cavity length $L = 60$ cm, and mean finesse $F = 1200$ for the P(7, $A_2$) and Q(7, $A_2$) dataset, and 1140 for the R(7, $A_2$) dataset. The figure of merit, defined as $\alpha_{min}(\tau/M)^{1/2}$, where $M = 2.5 \times 10^6$ is the number of spectral elements, is $3 \times 10^{-9}$ cm$^{-1}$ Hz$^{-1/2}$ per spectral element for the P(7, $A_2$)- and Q(7, $A_2$)-pumped spectra, and $7 \times 10^{-9}$ cm$^{-1}$ Hz$^{-1/2}$ per spectral element for the R(7, $A_2$)-pumped spectrum. These values are slightly worse than the figure of merit of $1.3 \times 10^{-9}$ cm$^{-1}$ Hz$^{-1/2}$ per spectral element reported in our previous work [47], because of the lower finesse of the new cavity.

### 3.4  Effective Hamiltonian

A novel methodology was recently proposed in Ref. [55] to construct effective Hamiltonians and dipole moment operators from PES and DMS. Unlike contact transformation based on perturbation theory, this approach obviates the need to make tedious algebraic calculations. Instead, we search for a numerical transformation $\mathbf{T}^{(J,C)}$ that bring $\mathbf{H}^{(J,C)}$ into a block-diagonal form $\mathbf{H}^{(J,C,P)}$ up to a polyad $P$. Here, $\mathbf{H}^{(J,C)}$ is the matrix representation of the complete nuclear-motion Hamiltonian for a given symmetry block $(J,C)$, and computed in a basis set $|\gamma, J, C\rangle$, where $\gamma$ denotes all other quantum numbers. In the case of methane, the polyad structure remains quite clear, even above 10,000 cm$^{-1}$, despite some energy overlap. We have shown that $\mathbf{T}^{(J,C)}$ and $\mathbf{H}^{(J,C,P)}$ are computed from selected variational eigenpairs. Finally, $\mathbf{H}^{(J,C,P)}$ is nothing but a matrix representation of an effective Hamiltonian $\tilde{H}(\tilde{t})$ in a basis $|\gamma, J, C, P\rangle$, where $\tilde{t}$ are parameters to be determined. Following the iterative procedure proposed in Ref. [55], these parameters are obtained by solving an overdetermined system of equations. In that work, a global effective Hamiltonian expended at order $\Omega_v + \Omega_r = 10$ was built up to the polyad $P = 6$ where $\Omega_v$ is the total vibrational degree ($\leq 10$ here) in the creation-annihilation operators $(a, a^+)$ and $\Omega_r$ ($\leq 6$) is the rotational degree in $(J_x, J_y, J_z)$. The resulting 14166 rovibrational *ab initio*-based effective parameters have been determined in only few minutes, and used to compute the energy levels of $P2$ and $P6$ up to $J = 9$. For line intensity calculation, $\boldsymbol{T}^{(J,C)}$ is used to transform the matrix of the laboratory-fixed frame dipole moment components computed in the same primitive basis set as the Hamiltonian. The unknown parameters $\tilde{\mu}$ of an effective dipole moment operator can be determined in a fashion similar to that used for computing the Hamiltonian's parameters. The line intensities of the $P6$ - $P2$ transitions were calculated from 990 dipole moment parameters. This new effective model was used in this work to unambiguously assign observed transitions up to $J = 9$.

## 4 Results

### 4.1 Ladder-type line parameters

We detected a total of 118 ladder-type OODR transitions: 35 in the P(7, $A_2$)-pumped spectrum, 34 in the Q(7, $A_2$)-pumped spectrum, 38 in the R(7, $A_2$)-pumped spectrum, and 11 in the Q(6, $F_2$)-pumped spectrum. The results for the $J'' = (7, A_2)$ pumped spectra are visualized by the black sticks in Figure 6, and compared to predicted transitions from the corresponding pumped levels (red sticks) obtained from the effective Hamiltonian [55], plotted in red and inverted for clarity. The Einstein A-coefficients of the predicted transitions have been arbitrarily scaled to match the experimental data. The figures are plotted in a common range of final state term values $E$ (upper x-scale), calculated as the sum of the experimental ladder-type transition wavenumbers, the pump transition frequencies known with kHz accuracy from measurements using sub-Doppler spectroscopy of the $\nu_3$ band from Refs. [17, 18] (listed in Table I), and lower state term value $E''$ calculated from empirically determined ground state molecular constants (obtained through a private communication with Hiroyuki Sasada). These term values are 293.1542822(8) cm$^{-1}$ for the $J'' = (7, A_2)$ state and 219.9149048(8) cm$^{-1}$ for the $J'' = (6, F_2)$ state.

The parameters of all detected lines are provided in Table VI – IX for the P(7, $A_2$) -, Q(7, $A_2$) -, R(7, $A_2$) - and Q(6, $F_2$) - pumped spectra, respectively. The uncertainties of line center frequencies retrieved by the fit were between 35 kHz and 3 MHz, depending on the SNR of the line. The total line center uncertainties were calculated as the quadrature sum of the fit uncertainty, the 330 kHz contribution from the sub-nominal resolution procedure, and the estimated pressure shift of 130 kHz. The uncertainties in the final state term values $E$ were calculated as the quadrature sum of our observed center frequency uncertainties and the uncertainties in the pumped energy levels. The latter we calculated from the transition frequency uncertainties reported in [17, 18] and the uncertainty in the pump lower state of 24 kHz. The Lorentzian widths of the probe transitions are listed with the fit uncertainty.

The integrated absorption coefficients of the Lorentzian components were corrected by the factors discussed in Section 3.1, and their uncertainties are the quadrature sum of the relative fit uncertainty, which varied between 0.5% and 30%, the 4% relative uncertainty of the finesse, as well as the uncertainty caused by the ellipticity of the probe. The latter was calculated similarly to what was done in the Supplementary Material of Ref. [47] and varied between 3 and 24%, depending on the change in rotational quantum numbers for the pump and probe transitions [59]. Relative intensity uncertainties due to probe ellipticity of ladder-type transitions belonging to the P, Q and R branches for each of the pump transitions are listed in Table II.

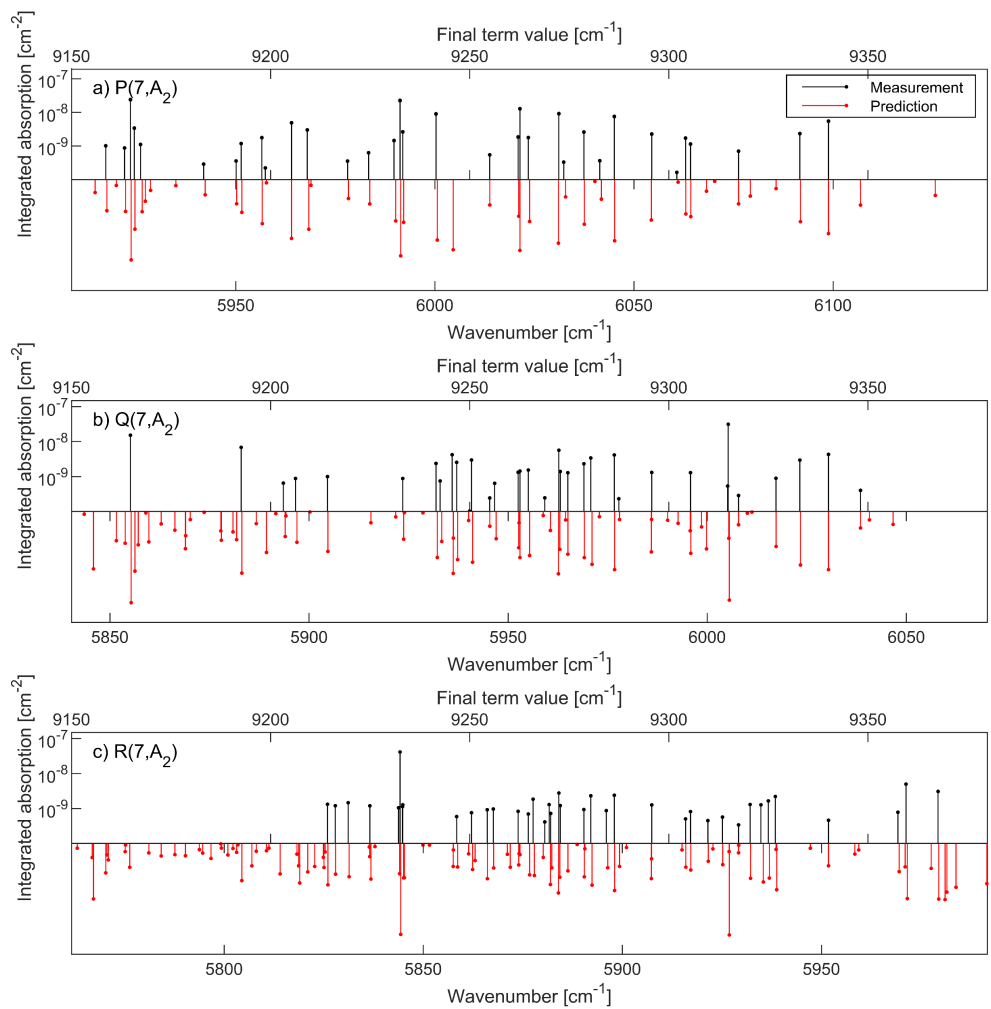

Figure 6. The ladder-type OODR transitions (black) detected when pumping the a) P(7, $A_2$), b) Q(7, $A_2$) and c) R(7, $A_2$) $\nu_3$ transitions compared to the predictions from the effective Hamiltonian (red), where the predicted Einstein $A$-coefficients are scaled by a common factor to match the experiment and plotted inverted. The integrated absorption coefficients is that of the ladder-type Lorentzian component only. The lower x-axis shows the transition wavenumber, while the upper x-axis shows the reached final state term value (common range in all panels).

Table II. The relative uncertainties in line intensities due to the ellipticity of the probe polarization for different combinations of pump and probe transitions. The rows indicate the four pump transitions, and columns 2-4 give the relative uncertainty for ladder-type transitions belonging to the P, Q, and R branches. The last column shows the relative uncertainty for the V-type transitions with the same rotational quantum number change as the pump transition.

| Pump | P - branch probe | Q - branch probe | R - branch probe | V – type probe |
|---|---|---|---|---|
| **P(7)** | 0.0534 | 0.0839 | 0.0335 | 0.1119 |
| **Q(7)** | 0.1464 | 0.2379 | 0.0980 | 0.2379 |
| **R(7)** | 0.1043 | 0.1739 | 0.0732 | 0.036 |
| **Q(6)** | 0.1514 | 0.2379 | 0.0952 | -- |

### 4.2  Combination differences

The final states common for transitions detected in the three $J'' = (7, A_2)$ - pumped spectra could be identified as clusters in observed final state term values, $E$, that agreed within $3 \times 10^{-4}$ cm$^{-1}$, which is 2 orders of magnitude lower than the minimum separation between distinct probed final states. We found 24 final states common between two or three measurements, and 58 of the 107 ladder-type transitions formed combination differences with at least one other line. Figure 7a)-c) shows three transitions from the R(7, $A_2$), Q(7, $A_2$) and P(7, $A_2$) - pumped spectra that reach the same final state, which implies that its rotational quantum number must be $J = 7$ (see Figure 2b). The upper panel shows the deviation between the final state term value obtained for these three transitions with respect to their weighted mean of 9286.312823(8) cm⁻¹, where the weights are the inverse of the squares of the uncertainties. The error bars are 1σ uncertainties of the individual final state term values. We found that for all but one of the common final state term values, the individual results agree within 3σ with their weighted mean (the one line deviating by slightly more than 3σ had a very low SNR). This shows that the wavenumber uncertainties are not underestimated.

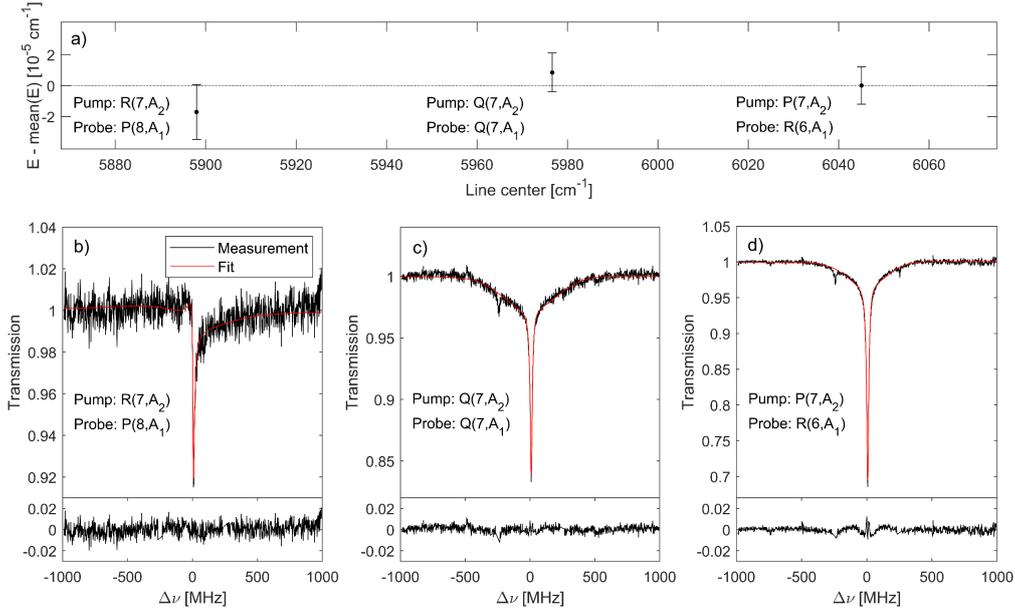

Figure 7. a) The agreement in the upper state term value reached by three different combinations of pump-probe transitions expressed as deviation from the weighted mean value of 9286.312823(8) cm⁻¹. The corresponding ladder-type OODR transitions, one for each pump transition, are shown in panels b)-d), together with the fits (red) and residuals in the bottom panels.

Finding a common final state in different spectra allows putting restrictions on the final state rotational quantum number $J$. From Figure 2b), final states with $J = 7$ can be reached in all three measurements, while upper states with $J = 6$ and $J = 8$ can be reached in two of them. A final state found in the P(7, $A_2$) and R(7, $A_2$)-pumped spectrum, but not Q(7, $A_2$)-pumped, also restricts its rotational quantum number to $J = 7$. The reason for not observing the state for the Q(7, $A_2$) pump transition could be either the line falling outside the measured wavelength range, being too weak, or concealed by a strong Doppler-broadened line absorbing all of the probe light. 12 of the 24 common final states could in this way be unambiguously assigned to $J = 7$. The remaining 12 final states were observed for P(7, $A_2$) and Q(7, $A_2$), or Q(7, $A_2$) and R(7, $A_2$) pump transitions, restricting their $J$ values to 6, 8 or 7, again allowing also for the possibility of a third member of the group not detected in the experiment.

### 4.3 Assignment and comparison to theory

We assigned the rotational quantum numbers of the final energy levels detected in the $J'' = (7, A_2)$ - pumped spectra using the combination differences described above and comparison to predictions from the effective Hamiltonian [55], shown in Figure 6. The correspondence between the experiment and predictions is immediately obvious. A few lines apparently missing from the measurements, such as the strong line at 5926 cm⁻¹ in the R(7, $A_2$) - pumped spectrum in Figure 6c), are due to overlap with a strong Doppler-broadened line. Most detected lines could easily be matched to predicted transitions.

In the majority of cases, observed lines were assigned to the closest predicted transition of comparable intensity, though occasionally common patterns in position and intensity were used for clusters of lines. We checked the validity of all assignments against the available combination differences, which were needed in a few cases to resolve ambiguous assignments in one measurement with the help of another line of the combination difference with a more undisputable match. All lines could be assigned, and these assignments are listed in Table VI - VIII.

For the Q(6, $F_2$) - pumped spectra, the assignment to the Hamiltonian was based only on comparing the observed and predicted transition wavenumbers and intensities, since no combination differences were observed for the final $F_2$ - symmetry states. The assignments of the Q(6, $F_2$) - pumped lien list are listed in Table IX.

In the four measurements, we detected a total of 84 final states and the term values of these are provided in Table X, where a weighted mean was taken in all cases where the same state was detected in more than one spectrum (with inverse of the square of the uncertainty as the weight). The table also lists the term values predicted from the Hamiltonian, and the corresponding rotational and vibrational assignments, and the counting numbers. The last column shows in which spectrum (spectra) the particular final state was detected.

We also assigned our line list using the TheoReTS/HITEMP [11] and ExoMol [13] line lists. Assignments to these could be carried out quite easily for the stronger lines. The choice was not always straightforward for weaker lines, though again the observed combination differences often allowed to narrow down the options. Additional assignments could be made by comparing the final state $J$-numbers to those of the corresponding assignments to the effective Hamiltonian, and we made sure that these assigned final $J$-numbers agreed between the three sets of assignments. In this way, all but three detected ladder-type lines (two for TheoReTS/HITEMP and one for ExoMol) could be assigned. For these lines, there either was no plausible candidate that agreed in $J$ with the Hamiltonian assignment, or there were several suitable candidate lines, not allowing for a conclusive match. The experimental line lists with TheoReTS/HITEMP and ExoMol assignments are provided in the Supplementary Material. The ExoMol assignments of the final states are listed in Table X.

Figure 8a) shows the difference between the observed and predicted transition wavenumbers from the Hamiltonian for each $J'' = (7, A_2)$ - pumped measurement. A similar comparison to the TheoReTS/HITEMP database is shown in Figure 9a), while Figure 10a) shows a comparison to the ExoMol database. The wavenumber agreement in Figure 8a) - Figure 10a) is clearly limited by the theoretical models, as the discrepancies are up to 4 orders of magnitude larger than the experimental uncertainties. The mean offsets from the models and the standard deviations of the discrepancies are given in Table III. The new Hamiltonian matches the measurement better than the other two line lists, with average wavenumber offset reduced by more than a factor of 4, and the scatter reduced by more than a factor of 3. The offsets are also negative in all cases, which is due to an apparent underestimation of the final state energies in all three sets of predictions.

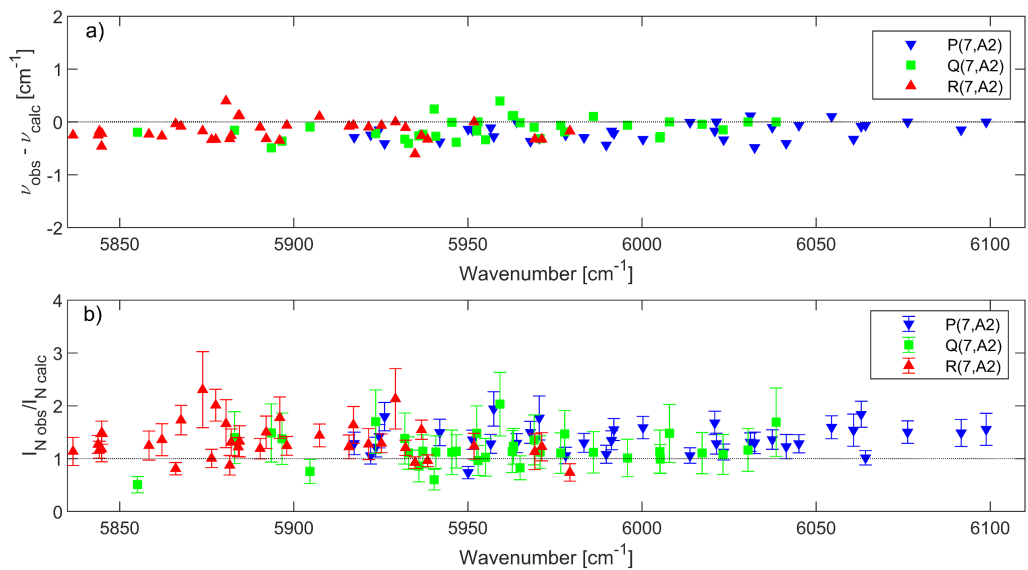

Figure 8. a) The difference between the observed center wavenumbers and those predicted by the effective Hamiltonian for ladder-type OODR lines detected in the measurements with three different pump transitions, as marked in the legend. b) The ratios of the normalized intensities obtained from the three measurements to those calculated based on data from the Hamiltonian and HITRAN. The error bars show 1σ uncertainties, and they are negligibly small in a).

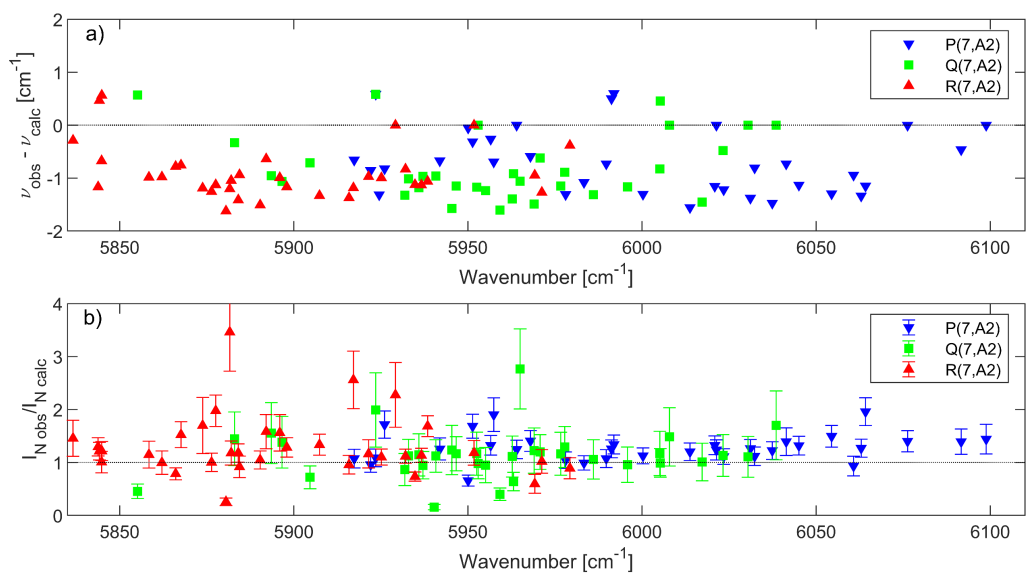

Figure 9. a) The difference between the observed center wavenumbers and those predicted by TheoReTS/HITEMP for ladder-type OODR lines detected in the measurements with three different pump transitions, as marked in the legend. b) The ratios of the normalized intensities obtained from the three measurements to those calculated based on data from the TheoReTS/HITEMP and HITRAN. The error bars show 1σ uncertainties, and they are negligibly small in a).

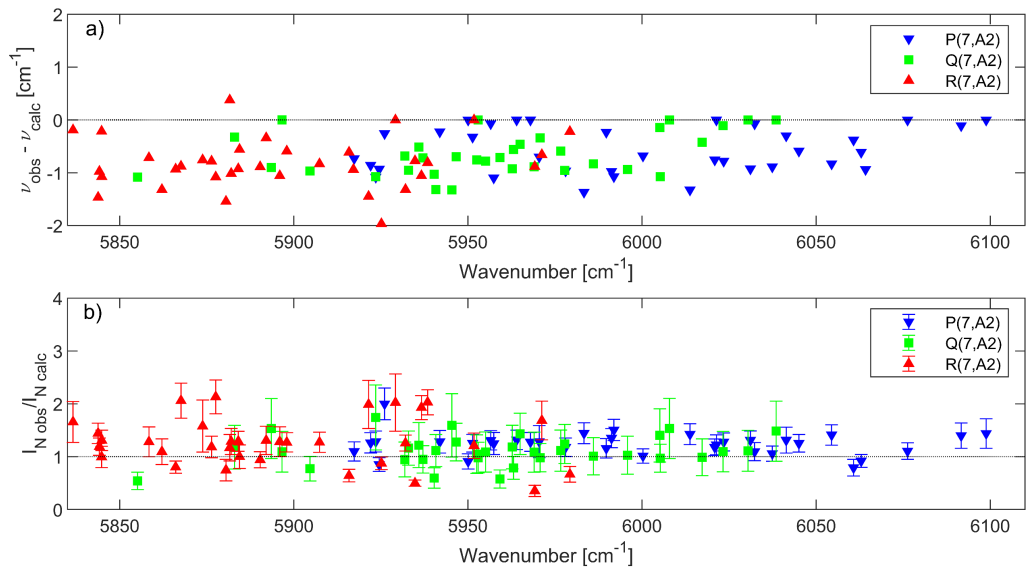

Figure 10. a) The difference between the observed center wavenumbers and those predicted by ExoMol for ladder-type OODR lines detected in the measurements with three different pump transitions, as marked in the legend. b) The ratios of the normalized intensities obtained from the three measurements to those calculated based on data from the ExoMol and HITRAN. The error bars show 1σ uncertainties, and they are negligibly small in a).

Table III. The mean values and standard deviations of the discrepancies in ladder-type center wavenumbers and normalized intensities between the experiment and the Hamiltonian, TheoReTS/HITEMP and ExoMol.

| Reference source | Center wavenumbers [cm⁻¹] | | Normalized intensities | |
|---|---|---|---|---|
| | **Mean offset** | **Standard deviation** | **Mean ratio** | **Standard deviation** |
| Hamiltonian | -0.16 | 0.18 | 1.2 | 0.32 |
| TheoReTS/HITEMP | -0.76 | 0.69 | 1.2 | 0.45 |
| ExoMol | -0.67 | 0.45 | 1.0 | 0.34 |

### 4.4 V-type line parameters

We detected 27 V-type transitions in the four spectra starting from the ground vibrational levels with $J'' = (7, A_2)$ and $(6, F_2)$. Their transition wavenumbers and assignments based on the effective Hamiltonian are listed in Table XI. The wavenumber uncertainties are also here calculated as the quadrature sum of the fit uncertainties and the contributions from the sub-nominal resolution procedure and the pressure shift. Many of the V-type transitions from the $J'' = (7, A_2)$ state appeared in at least 2 spectra, and the wavenumber listed in the table is the weighted mean of the results from fits to the different spectra.

These transitions belong to different combination bands and can also be found in HITRAN2020 [57], ExoMol [13], and in the WKLMC line list of Nikitin *et al.* [29]. In all three cases, the assignment was straightforward to the closest predicted strong line. Tables of comparisons to HITRAN2020, ExoMol, and the WKLMC line list can be found in the Supplementary Material. For some lines, the vibrational assignment differs between the Hamiltonian and ExoMol, because of mixing between the vibrational states. In HITRAN, the vibrational assignment is missing for most of the lines, and for some even the rotational assignment is missing. The WKLMC line list does not contain the vibrational assignment.

Figure 11 shows a comparison of the observed center wavenumbers to the four line lists and Table IV summarizes the mean and standard deviations of the discrepancies between them and the experimental values. Here the line list from the effective Hamiltonian, which is the only one not based on empirical data, displays the largest scatter (though significantly smaller than for the ladder-type predictions).

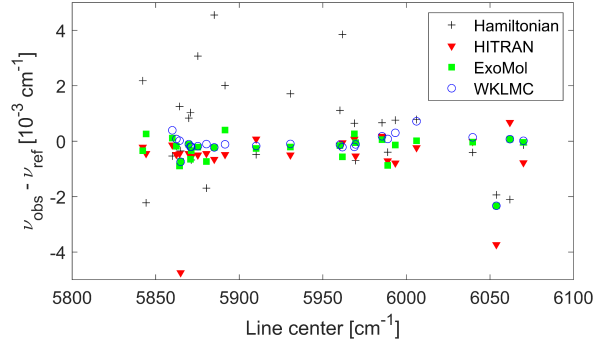

Figure 11. A comparison of the observed center wavenumbers of the V-type features to the effective Hamiltonian (black crosses), HITRAN (red triangles), ExoMol (green squares), and WKLMC line list (blue circles). The experimental uncertainties are negligible compared to the scatter.

Table IV. The mean offset between the experimental center frequencies of the V-type lines and those from the Hamiltonian, HITRAN2020, ExoMol and WKLMC, as well as the standard deviation of the scatter.

| Reference | Mean offset [cm$^{-1}$] | Standard deviation [cm$^{-1}$] |
|---|---|---|
| Hamiltonian | $4.7 \cdot 10^{-4}$ | $1.7 \cdot 10^{-3}$ |
| HITRAN | $-6.2 \cdot 10^{-4}$ | $1.1 \cdot 10^{-3}$ |
| ExoMol | $-2.9 \cdot 10^{-4}$ | $5.3 \cdot 10^{-4}$ |
| WKLMC | $-1.3 \cdot 10^{-4}$ | $5.3 \cdot 10^{-4}$ |

Table V shows the center wavenumbers and intensities of the three V-type resonances selected in each $J'' = (7, A_2)$ - pumped spectrum for normalization of the ladder-type intensities, together with their Hamiltonian assignment. The Einstein $A$-coefficients are taken from HITRAN2020 [57], and they differ from those from the Hamiltonian by <1.5%. The integrated absorption coefficients of the V-type features refer to their Lorentzian components only, and their uncertainties are the combination of fit, finesse and polarization-induced uncertainties (see Table II).

Table V. Parameters of the three V-type transitions from the $\nu_3$ $J'' = (7,A_2)$ state used for quantification of the ladder-type line intensities. Column 1: Pump transition of the spectrum in which the V-type appears. Columns 2 and 3: The rotational and vibrational assignment of the transitions displaying the V-type features (from the effective Hamiltonian). Column 4 and 5: The transition wavenumber and integrated absorption coefficient of the Lorentzian component of the V-type feature. Column 6: Einstein $A$-coefficient from HITRAN2020. Column 7: The upper state degeneracy of the transitions.

| Pump transition | V-type probe transition | | | | | |
|---|---|---|---|---|---|---|
| $\nu_3$ band | Rotational assignment | Vibrational assignment | Observed wavenumber [cm$^{-1}$] | V-type integrated absorption [$10^{-9}$ cm$^{-2}$] | Einstein $A$-coefficient (HITRAN) [s$^{-1}$] | Upper state deg. |
| P(7, $A_2$) | P(7, $A_2$) | $2\nu_2 + \nu_3$ | 5993.38190(4) | 0.59(7) | 0.01653 | 65 |
| Q(7, $A_2$) | Q(7, $A_2$) | $2\nu_3$ | 6039.65774(7) | 1.4(4) | 0.03194 | 75 |
| R(7, $A_2$) | R(7, $A_2$) | $\nu_2 + \nu_3 + \nu_4$ | 5910.12483(6) | 5.1(4) | 0.0973 | 85 |

### 4.5 Normalized line intensities

We calculated the normalized experimental intensities, $I_{N\,obs}$, of the OODR ladder-type transitions detected in the $J'' = (7, A_2)$ - pumped spectra as the ratios of the integrated absorption of the Lorentzian components of the ladder-type transitions (from Table VI - VIII) and the V-type transition from the same spectrum (from Table V). These normalized intensities are compared to theoretical predictions from the Hamiltonian, TheoReTS/HITEMP and ExoMol in Figure 8b) - Figure 10b), respectively. The predicted normalized intensities were calculated as $I_{N\,calc} = (A_{LT}\, g'_{LT}\, / \, \nu'_{LT}) \, / \, (A_{VT}\, g'_{VT}\, / \, \nu'_{VT})$ [46], where $A_k$ are the Einstein $A$-coefficients, $g'_k$ the upper state degeneracies, and $\nu_k$ the transition wavenumbers of the ladder-type ($k$ = LT) and V-type ($k$ = VT) transitions, respectively. The Einstein $A$-coefficients for the ladder-type transitions were taken from the effective Hamiltonian, TheoReTS/HITEMP and ExoMol, respectively, while for the ground state transitions displaying the V-type transition they were taken from HITRAN (see Table V). The upper state degeneracies were calculated as $g_k' = (2J' + 1)s$, where $s$ is the spin factor equal to 5 for $A_1$ and $A_2$, symmetry states, and 3 for $F_1$ and $F_2$ symmetry states.

Figure 8b) - Figure 10b) show the ratios $I_{N\,obs}/I_{N\,calc}$ of the observed normalized intensities to the predicted counterparts calculated from the Hamiltonian, TheoReTS/HITEMP and ExoMol, respectively. The error bars are the combined uncertainty of the integrated absorption coefficients of the Lorentzian components of the ladder-type and the V-type transitions. The mean values of the ratios and the standard deviations of the scatter are given in Table III. For TheoReTS/HITEMP we excluded an outlier deviating by a factor of 10, out of scale in Figure 9b), when calculating the numbers. There is no other plausible assignment for this line. For ExoMol, two outliers were excluded, occurring off the y-scale range around 5917 cm$^{-1}$ in the R(7, $A_2$)-pumped spectrum and around 6064 cm$^{-1}$ in the P(7, $A_2$)-pumped spectrum. These two lines belong to the same combination difference triplet and we see no alternative assignment. We note that although the relative uncertainty in the V-type integrated absorption is between 8% and 30% for the three measurements, inaccuracies in these values would only produce offsets in Figure 8b) - Figure 10b) while the scatter would remain the same.

## 5   Conclusion

We used cavity-enhanced comb-based OODR spectroscopy to expand the experimental analysis of the $P6 - P2$ region of CH$_4$ to rotational levels higher than $J = 4$. The broad coverage of the comb probe allowed detection of 118 previously unobserved hot-band transitions from four ro-vibrational levels in the $\nu_3$ band reaching final states with rotational quantum numbers $J = 5$ to 9 in the 9070-9370 cm$^{-1}$ range. We assigned the final state $J$ values of these transitions based on combination differences and theoretical predictions from a new effective Hamiltonian. We found that the experimental line positions agree better with the new Hamiltonian predictions than with TheoReTS/HITEMP, for which larger deviations may occur for high $J$ values because of poorer convergence of the variational calculation. Both the TheoReTS/HITEMP and ExoMol predictions show a mean systematic offset of -0.7 cm$^{-1}$ with respect to the experiment. For ExoMol, this is a significant improvement compared to its earlier versions, where the discrepancies were of the order of a few wavenumbers [46]. The agreement between the normalized line intensities was similar for the three sets of predictions, with ratios of around 1-1.2 and a scatter around 0.3. We also reported 27 line positions of ground state transitions to the 2$\nu_3$ region and found a better agreement with the ExoMol and WKLMC line lists than with the new Hamiltonian.

This work allowed to validate for the first time the accuracy of the *ab initio* effective Hamiltonian used to assign lines of a very crowded methane polyad. The data provided by earlier and present OODR measurements will help improve the parameterization of the effective Hamiltonian for the triacontad. The measured spectra also contain a large number of Doppler-broadened 4-level OODR transitions induced by collisional redistribution of the population transferred by the pump. Analysis of those transitions is under way and will be reported in subsequent work.

## Supplementary material

The supplementary material contains two tables and is available from the corresponding author. The first table lists line positions, upper state term values and absorption coefficients of the ladder-type transitions compared to line positions, upper state term values and Einstein A coefficients from TheoReTS/HITEMP and ExoMol. The second table lists the experimental positions of the V-type transitions together with line positions, line intensities, upper state vibrational assignments and counting numbers (when available) from the effective Hamiltonian, ExoMol, HITRAN2020 and the WKLMC line list.

## Acknowledgments

The authors thank Hiroyuki Sasada for providing the ground state term value from his unpublished work. This project is supported by the Knut and Alice Wallenberg Foundation (grant: KAW 2020.0303), and the Swedish Research Council (grant: 2020-00238). K.K.L. acknowledges funding from the U.S. National Science Foundation (grant: CHE-2108458). M.R. acknowledges support from the French ANR TEMMEX project (grant: 21-CE30-0053-01). L.R acknowledges support from the French National Research Agency (grant: ANR-19-CE30-0038). G.S. acknowledges support from the Foundation for Polish Science (grant: POIR.04.04.00-00-434D/17-00).

## Author declarations

The authors have no conflicts to disclose.

Adrian Hjältén: Investigation (equal), Formal Analysis (lead), Data Curation (equal), Validation (lead), Methodology (equal), Visualization (equal), Writing/Original Draft Preparation (equal)

Vinicius Silva de Oliveira: Investigation (equal), Data Curation (equal), Formal Analysis (supporting), Visualization (equal), Methodology (equal), Writing/Original Draft Preparation (equal)

Isak Silander: Investigation (equal), Formal Analysis (supporting), Methodology (equal)

Andrea Rosina: Investigation (equal), Formal Analysis (supporting), Writing/Review & Editing (supporting)

Michael Rey: Methodology (equal), Formal Analysis (supporting), Validation (supporting), Writing/Original Draft Preparation (equal)

Lucile Rutkowski: Validation (supporting), Writing/Review & Editing (supporting)

Grzegorz Soboń: Resources (supporting), Writing/Review & Editing (supporting)

Kevin K. Lehmann: Conceptualization (equal), Formal Analysis (supporting), Validation (supporting), Writing/Review & Editing (supporting)

Aleksandra Foltynowicz: Conceptualization (equal), Funding Acquisition (lead), Project Administration (lead), Supervision (lead), Resources (lead), Writing/Review & Editing (lead)

## Data availability statement

The spectra that support the findings of this study are available from the corresponding author upon reasonable request.

# Appendices

## *A. Baseline correction and interleaving of Doppler-broadened spectra*

Before interleaving the background and OODR spectra, at each $f_{rep}$ step, we removed the baseline originating from the comb envelope and etalon effects from the transmission spectra. To find the baseline we applied a custom non-linear fitting routine to the spectra in the entire acquired spectral range. The Doppler-broadened lines were modeled using the cavity transmission function [61] with a Gaussian line shape and line parameters (positions and intensities) fixed to the values from the HITRAN2020 database [57]. The methane fraction in the model was fixed to 1, the cavity finesse was fitted as a global constant value, and the comb-cavity-offset phase was modeled as a cubic Bézier curve, with the control points as fitting parameters. The baseline was found using the cepstral analysis [65], which operates on the inverse Fourier transform of the negative natural logarithm of the spectrum (cepstrum), where the different contributions (OODR signal, Doppler-broadened lines, baseline) are summed, instead of multiplied. The baseline was set to contain structures with periods of the order of tens to hundreds of GHz or longer. We note that this method of baseline retrieval is insensitive to inaccuracies in the shape and intensity of the Doppler-broadened lines, caused by inaccuracies in the comb-cavity-offset and the finesse values, because the removed baseline spectral features are much wider than these lines. To normalize the transmission spectra, we divided the spectra at each $f_{rep}$ step by the corresponding baseline, then interleaved the baseline-corrected spectra and averaged all acquisition series.

## *B. Baseline correction and interleaving of spectra for OODR line fitting*

To remove the slowly varying baseline remaining in the normalized OODR spectra at each $f_{rep}$ step, we first calculated a rolling average over 15-GHz-wide segments. Next, we found points in the original spectrum that deviated by more than one standard deviation from the mean of the rolling average over the segment. We replaced these points and their two nearest neighbors with this mean value. This smoothed out the noisy parts where Doppler-broadened absorption was strong, and removed the sub-Doppler and Doppler-broadened absorption lines, because their HWHM linewidth is ~8 and ~275 MHz, respectively, so that at each $f_{rep}$ step only one comb mode is absorbed by a given sub-Doppler line, and 3-5 comb lines are absorbed by each Doppler-broadened line. The baseline was removed from the cepstrum [65] of the smoothed spectrum and contained components corresponding to spectral features with periods larger than 6.4 GHz (0.21 cm$^{-1}$). Afterwards, we divided each normalized spectrum by its corresponding baseline, interleaved the baseline-corrected spectra to 2 MHz sampling point spacing, and averaged all acquisition series. These interleaved baseline-corrected normalized spectra contained the sub-Doppler OODR transitions and the Doppler-broadened transitions not cancelled by the normalization process.

Before fitting the sub-Doppler OODR lines, we removed the remaining Doppler-broadened lines appearing within ±2.5 GHz from the centers of the OODR lines (determined from the peak-finding routine). To find the centers of the Doppler-broadened lines, we convolved the spectral segment with a Gaussian line shape with HWHM of 280 MHz to suppress the sub-Doppler peaks. We then found the center frequencies using the MATLAB *findpeaks* function. The Doppler-broadened features were weak, so we modelled them as exponentials of Gaussian line shapes (i.e. as in the Lambert-Beer law) with the Doppler HWHM fixed to the room temperature value (~275 MHz). We fitted this model to each line in the ±2.5 GHz wide spectral segments with intensity as the only fitting parameter, and finally cancelled them by dividing the segment by the fitted model.

## C. Tables with line lists and energy levels

Table VI. Parameters of the lines detected in the P(7, $A_2$)-pumped measurement, i.e. starting from the level with $J' = (6, A_1)$ and term value 293.1542822(8) + 2948.10794477(8) = 3241.2622270(8) cm$^{-1}$. Column 1 lists the measured center wavenumber. Column 2 shows the final state term value calculated as described in Section 4.1. Column 3 states the integrated absorption coefficient of the Lorentzian component of the probe transition. Column 4 shows the half width at half maximum of the Lorentzian component of the probe transition. Columns 5 and 6 show the predicted probe transition wavenumbers from the effective Hamiltonian and its difference with respect to the observed wavenumber. Column 7 shows the Einstein A coefficient from the Hamiltonian. Column 8 shows the final state $J$ number and symmetry.

| 1 | 2 | 3 | 4 | 5 | 6 | 7 | 8 |
|---|---|---|---|---|---|---|---|
| Probe transition wavenumber [cm$^{-1}$] | Final state term value [cm$^{-1}$] | Probe transition integrated absorption coefficient [$10^{-9}$ cm$^{-2}$] | Probe transition width [MHz] | Hamiltonian transition wavenumber [cm$^{-1}$] | Obs. − pred. wavenum [cm$^{-1}$] | Hamiltonian Einstein A coeff. [s$^{-1}$] | Final state $J$ number |
| 5917.34032(5) | 9158.60255(5) | 1.0(1) | 9(1) | 5917.63107 | -0.29075 | 0.01869 | 7A2 |
| 5922.06831(3) | 9163.33053(3) | 0.88(8) | 7.7(9) | 5922.3135 | -0.24519 | 0.01967 | 7A2 |
| 5923.53428(1) | 9164.79651(1) | 24(2) | 9.4(2) | 5923.72773 | -0.19344 | 0.5427 | 6A2 |
| 5924.51887(1) | 9165.78110(1) | 3.4(3) | 8.3(3) | 5924.66476 | -0.14589 | 0.06603 | 6A2 |
| 5926.06992(5) | 9167.33214(5) | 1.1(1) | 13(1) | 5926.47905 | -0.40914 | 0.02007 | 5A2 |
| 5941.92108(4) | 9183.18331(4) | 0.29(3) | 8(1) | 5942.29973 | -0.37866 | 0.006292 | 5A2 |
| 5950.06485(3) | 9191.32708(3) | 0.36(4) | 7.4(8) | 5950.20263 | -0.13778 | 0.01165 | 7A2 |
| 5951.34689(1) | 9192.60912(1) | 1.19(7) | 8.1(3) | 5951.50214 | -0.15524 | 0.02096 | 7A2 |
| 5956.54279(1) | 9197.80501(1) | 1.8(1) | 8.4(1) | 5956.65205 | -0.10926 | 0.04545 | 5A2 |
| 5957.41611(5) | 9198.67833(5) | 0.22(3) | 10(1) | 5957.69306 | -0.27695 | 0.002766 | 7A2 |
| 5963.98956(1) | 9205.25178(1) | 5.0(3) | 8.81(8) | 5963.9899 | -0.00034 | 0.1251 | 5A2 |
| 5967.95407(1) | 9209.21630(1) | 3.0(2) | 8.34(9) | 5968.31969 | -0.36561 | 0.06641 | 5A2 |
| 5970.47685(7) | 9211.73907(7) | 0.09(2) | 8(2) | 5970.78363 | -0.30678 | 0.001220 | 7A2 |
| 5978.05358(3) | 9219.31581(3) | 0.36(3) | 7.8(7) | 5978.2978 | -0.24422 | 0.008134 | 7A2 |
| 5983.33610(2) | 9224.59833(2) | 0.63(4) | 9.1(4) | 5983.63022 | -0.29413 | 0.01176 | 7A2 |
| 5989.72932(2) | 9230.99155(2) | 1.5(1) | 7.2(4) | 5990.16643 | -0.43710 | 0.03772 | 6A2 |
| 5991.23448(1) | 9232.49671(1) | 23(1) | 10.04(8) | 5991.40771 | -0.17323 | 0.4109 | 7A2 |
| 5991.92035(1) | 9233.18258(1) | 2.7(1) | 8.1(1) | 5992.13763 | -0.21727 | 0.04146 | 7A2 |
| 6000.27330(1) | 9241.53553(1) | 9.1(5) | 8.8(1) | 6000.60133 | -0.32803 | 0.1393 | 7A2 |
| 6013.77138(2) | 9255.03361(2) | 0.55(4) | 6.9(4) | 6013.77885 | -0.00747 | 0.01264 | 7A2 |
| 6020.88677(1) | 9262.14900(1) | 1.9(1) | 8.2(1) | 6021.05712 | -0.17034 | 0.02724 | 7A2 |
| 6021.35259(1) | 9262.61482(1) | 13(1) | 9.08(6) | 6021.3532 | -0.00061 | 0.2841 | 6A2 |
| 6023.44714(1) | 9264.70937(1) | 1.8(1) | 8.3(1) | 6023.78337 | -0.33623 | 0.03907 | 7A2 |
| 6031.12164(1) | 9272.38387(1) | 9.3(5) | 9.01(6) | 6031.00775 | 0.11390 | 0.1731 | 7A2 |
| 6032.33755(2) | 9273.59978(2) | 0.33(3) | 7.1(3) | 6032.82214 | -0.48459 | 0.007255 | 6A2 |
| 6037.40452(1) | 9278.66675(1) | 2.6(2) | 7.7(2) | 6037.506 | -0.10148 | 0.04723 | 7A2 |
| 6041.39753(4) | 9282.65976(4) | 0.37(5) | 9(1) | 6041.80272 | -0.40519 | 0.008470 | 6A2 |
| 6045.05060(1) | 9286.31282(1) | 7.6(4) | 8.47(8) | 6045.11464 | -0.06405 | 0.1463 | 7A2 |
| 6054.44744(2) | 9295.70966(2) | 2.3(1) | 8.4(3) | 6054.34296 | 0.10448 | 0.03539 | 7A2 |
| 6060.74725(6) | 9302.00947(6) | 0.17(3) | 9(2) | 6061.0736 | -0.32635 | 0.002654 | 7A2 |
| 6062.93171(1) | 9304.19394(1) | 1.74(9) | 8.1(2) | 6063.01424 | -0.08253 | 0.02333 | 7A2 |
| 6064.18833(1) | 9305.45056(1) | 1.15(6) | 8.8(2) | 6064.25122 | -0.06289 | 0.02807 | 7A2 |
| 6076.25377(3) | 9317.51600(3) | 0.71(5) | 10.4(8) | 6076.2549 | -0.00113 | 0.01169 | 7A2 |
| 6091.64743(3) | 9332.90966(3) | 2.4(3) | 9.4(9) | 6091.79699 | -0.14956 | 0.03966 | 7A2 |
| 6098.82913(3) | 9340.09136(3) | 5.5(8) | 8.2(8) | 6098.8302 | -0.00107 | 0.08922 | 7A2 |

Table VII. Parameters of the lines detected in the Q(7, $A_2$)-pumped measurement, i.e. starting from the level with $J' = (7, A_1)$ and term value $293.1542822(8) + 3016.49766637(6) = 3309.6519486(8)$ cm$^{-1}$. Column 1 lists the measured center wavenumber. Column 2 shows the final state term value calculated as described in Section 4.1. Column 3 states the integrated absorption coefficient of the Lorentzian component of the probe transition. Column 4 shows the half width at half maximum of the Lorentzian component of the probe transition. Columns 5 and 6 show the predicted probe transition wavenumbers from the effective Hamiltonian and its difference with respect to the observed wavenumber. Column 7 shows the Einstein A coefficient from the Hamiltonian. Column 8 shows the final state $J$ number and symmetry.

| 1 | 2 | 3 | 4 | 5 | 6 | 7 | 8 |
|---|---|---|---|---|---|---|---|
| Probe transition wavenumber [cm$^{-1}$] | Final state term value [cm$^{-1}$] | Probe transition integrated absorption coefficient [$10^{-9}$ cm$^2$] | Probe transition width [MHz] | Hamiltonian transition wavenumber [cm$^{-1}$] | Obs. – pred. wavenumber [cm$^{-1}$] | Hamiltonian Einstein A coeff. [s$^{-1}$] | Final state $J$ number |
| 5855.14456(2) | 9164.79651(2) | 15(3) | 6.7(6) | 5855.33844 | -0.19387 | 0.7320 | 6A2 |
| 5882.95713(2) | 9192.60908(2) | 7(2) | 10.7(6) | 5883.11285 | -0.15571 | 0.1041 | 7A2 |
| 5893.53746(6) | 9203.18941(6) | 0.7(2) | 11(2) | 5894.02595 | -0.48849 | 0.009348 | 7A2 |
| 5896.60171(2) | 9206.25366(2) | 0.9(2) | 8.2(6) | 5896.96541 | -0.36369 | 0.01366 | 7A2 |
| 5904.63787(4) | 9214.28982(4) | 1.0(2) | 6(1) | 5904.73519 | -0.09732 | 0.02489 | 8A2 |
| 5923.53057(4) | 9233.18252(4) | 0.9(2) | 14(1) | 5923.74834 | -0.21777 | 0.01116 | 7A2 |
| 5931.88359(1) | 9241.53554(1) | 2.4(6) | 9.1(2) | 5932.21204 | -0.32845 | 0.03714 | 7A2 |
| 5932.92915(2) | 9242.58110(2) | 0.75(9) | 9.7(6) | 5933.3354 | -0.40625 | 0.01293 | 8A2 |
| 5935.93801(1) | 9245.58996(1) | 4(1) | 9.1(2) | 5936.20074 | -0.26273 | 0.1060 | 7A2 |
| 5937.07368(1) | 9246.72562(1) | 2.6(3) | 8.9(2) | 5937.3056 | -0.23192 | 0.04285 | 8A2 |
| 5940.31074(5) | 9249.96268(5) | 0.10(2) | 5(1) | 5940.06631 | 0.24442 | 0.003248 | 8A2 |
| 5940.80148(1) | 9250.45343(1) | 3.0(3) | 9.6(2) | 5941.07062 | -0.26914 | 0.05067 | 8A2 |
| 5945.38167(7) | 9255.03362(7) | 0.24(7) | 10(2) | 5945.38956 | -0.00790 | 0.004732 | 7A2 |
| 5946.61705(3) | 9256.26899(3) | 0.64(7) | 10.5(7) | 5947.00255 | -0.38550 | 0.01080 | 8A2 |
| 5952.49709(1) | 9262.14904(1) | 1.3(3) | 9.0(3) | 5952.66783 | -0.17074 | 0.01953 | 7A2 |
| 5952.96290(2) | 9262.61485(2) | 1.4(2) | 8.9(3) | 5952.96391 | -0.00101 | 0.03732 | 6A2 |
| 5955.05742(1) | 9264.70937(1) | 1.5(4) | 8.5(2) | 5955.39408 | -0.33666 | 0.03263 | 7A2 |
| 5959.19852(6) | 9268.85047(6) | 0.25(4) | 11(2) | 5958.80211 | 0.39641 | 0.002335 | 8A2 |
| 5962.73193(1) | 9272.38388(1) | 6(1) | 9.2(2) | 5962.61846 | 0.11347 | 0.1092 | 7A2 |
| 5963.09063(1) | 9272.74258(1) | 1.4(1) | 8.3(2) | 5962.96945 | 0.12118 | 0.02175 | 8A2 |
| 5964.98937(2) | 9274.64132(2) | 1.3(1) | 10.1(3) | 5965.00502 | -0.01565 | 0.02995 | 8A2 |
| 5969.01482(1) | 9278.66677(1) | 2.3(6) | 9.2(2) | 5969.11671 | -0.10189 | 0.03757 | 7A2 |
| 5970.74800(1) | 9280.39994(1) | 3.4(4) | 9.0(1) | 5971.06433 | -0.31634 | 0.05853 | 8A2 |
| 5976.66088(1) | 9286.31283(1) | 4(1) | 9.5(2) | 5976.72535 | -0.06447 | 0.08263 | 7A2 |
| 5977.79464(5) | 9287.44658(5) | 0.23(4) | 8(1) | 5977.98056 | -0.18593 | 0.003060 | 8A2 |
| 5986.05772(2) | 9295.70967(2) | 1.3(3) | 7.5(4) | 5985.95367 | 0.10406 | 0.02578 | 7A2 |
| 5995.79857(2) | 9305.45052(2) | 1.3(3) | 7.4(3) | 5995.86193 | -0.06336 | 0.02819 | 7A2 |
| 6005.11860(5) | 9314.77055(5) | 0.5(1) | 11(1) | 6005.4229 | -0.30429 | 0.01046 | 7A2 |
| 6005.26745(1) | 9314.91940(1) | 31(3) | 12.4(1) | 6005.5437 | -0.27625 | 0.6168 | 8A2 |
| 6007.86408(6) | 9317.51603(6) | 0.29(8) | 8(2) | 6007.86561 | -0.00153 | 0.004305 | 7A2 |
| 6017.24262(3) | 9326.89457(3) | 0.9(2) | 9.0(7) | 6017.28763 | -0.04501 | 0.01802 | 7A2 |
| 6023.25772(1) | 9332.90967(1) | 3.0(7) | 9.1(2) | 6023.4077 | -0.14998 | 0.06108 | 7A2 |
| 6030.43943(2) | 9340.09138(2) | 4(1) | 9.1(3) | 6030.44091 | -0.00148 | 0.08282 | 7A2 |
| 6038.48580(6) | 9348.13775(6) | 0.4(1) | 7(2) | 6038.48701 | -0.00121 | 0.005328 | 7A2 |

Jun 12, 2024

Table VIII. Parameters of the lines detected in the R(7, $A_2$)-pumped measurement, i.e. starting from the level with $\mathcal{J} = (8, A_1)$ and term value 293.1542822(8) + 3095.17923673(13) = 3388.3335189(8) cm⁻¹. Column 1 lists the measured center wavenumber. Column 2 shows the final state term value calculated as described in Section 4.1. Column 3 states the integrated absorption coefficient of the Lorentzian component of the probe transition. Column 4 shows the half width at half maximum of the Lorentzian component of the probe transition. Columns 5 and 6 show the predicted probe transition wavenumbers from the effective Hamiltonian and its difference with respect to the observed wavenumber. Column 7 shows the Einstein A coefficient from the Hamiltonian. Column 8 shows the final state $J$ number and symmetry.

| 1 | 2 | 3 | 4 | 5 | 6 | 7 | 8 |
|---|---|---|---|---|---|---|---|
| Probe transition wavenumber [cm⁻¹] | Final state term value [cm⁻¹] | Probe transition integrated absorption coefficient [10⁻⁹ cm⁻²] | Probe transition width [MHz] | Hamiltonian transition wavenumber [cm⁻¹] | Obs. − pred. wavenumber [cm⁻¹] | Hamiltonian Einstein A coeff. [s⁻¹] | Final state $J$ number |
| 5825.95627(4) | 9214.28979(4) | 1.3(3) | 6(1) | 5826.05374 | -0.09747 | 0.02852 | 8A2 |
| 5827.91737(7) | 9216.25089(7) | 1.2(3) | 9(2) | 5827.99405 | -0.07668 | 0.01410 | 8A2 |
| 5831.14867(4) | 9219.48219(4) | 1.5(3) | 8(1) | 5831.38085 | -0.23218 | 0.01673 | 8A2 |
| 5836.59606(5) | 9224.92958(5) | 1.2(3) | 9(1) | 5836.84691 | -0.25085 | 0.01949 | 8A2 |
| 5843.80206(3) | 9232.13558(3) | 1.0(1) | 7.6(8) | 5844.05267 | -0.25061 | 0.01374 | 9A2 |
| 5844.16319(1) | 9232.49671(1) | 41(5) | 11.6(2) | 5844.33697 | -0.17378 | 0.7468 | 7A2 |
| 5844.79566(3) | 9233.12918(3) | 1.1(2) | 8.1(7) | 5845.25757 | -0.46191 | 0.01784 | 8A2 |
| 5844.84907(2) | 9233.18259(2) | 1.3(2) | 7.3(6) | 5845.06689 | -0.21782 | 0.01808 | 7A2 |
| 5858.39215(5) | 9246.72567(5) | 0.6(1) | 8(1) | 5858.62415 | -0.23200 | 0.008799 | 8A2 |
| 5862.11988(4) | 9250.45340(4) | 0.8(2) | 7(1) | 5862.38917 | -0.26929 | 0.01042 | 8A2 |
| 5866.06941(3) | 9254.40293(3) | 0.9(1) | 6.4(7) | 5866.10506 | -0.03565 | 0.01906 | 9A2 |
| 5867.60058(4) | 9255.93410(4) | 1.0(1) | 8(1) | 5867.68276 | -0.08218 | 0.009387 | 9A2 |
| 5873.81540(8) | 9262.14892(8) | 0.8(3) | 10(2) | 5873.98638 | -0.17097 | 0.007675 | 7A2 |
| 5876.37597(6) | 9264.70948(6) | 0.7(1) | 10(2) | 5876.71263 | -0.33666 | 0.01479 | 7A2 |
| 5877.57335(3) | 9265.90687(3) | 1.8(2) | 8.7(8) | 5877.9048 | -0.33145 | 0.01545 | 9A2 |
| 5880.51699(6) | 9268.85051(6) | 0.4(1) | 7(2) | 5880.12066 | 0.39633 | 0.004683 | 8A2 |
| 5881.60534(3) | 9269.93886(3) | 1.3(3) | 8.0(8) | 5881.92613 | -0.32079 | 0.02790 | 8A2 |
| 5882.01487(7) | 9270.34839(7) | 0.7(1) | 10(2) | 5882.27403 | -0.25915 | 0.009322 | 9A2 |
| 5884.05033(2) | 9272.38384(2) | 2.8(3) | 7.5(4) | 5883.937 | 0.11332 | 0.04863 | 7A2 |
| 5884.40898(3) | 9272.74250(3) | 1.2(3) | 8.1(9) | 5884.288 | 0.12099 | 0.01724 | 8A2 |
| 5890.33320(3) | 9278.66671(3) | 0.9(1) | 6.8(8) | 5890.43526 | -0.10206 | 0.01680 | 7A2 |
| 5892.06640(2) | 9280.39992(2) | 2.3(4) | 7.2(5) | 5892.38288 | -0.31648 | 0.02910 | 8A2 |
| 5895.96633(3) | 9284.29985(3) | 0.9(2) | 7.2(9) | 5896.31626 | -0.34992 | 0.009271 | 8A2 |
| 5897.97929(2) | 9286.31281(2) | 2.4(3) | 7.9(4) | 5898.0439 | -0.06461 | 0.04146 | 7A2 |
| 5907.37617(3) | 9295.70969(3) | 1.3(2) | 9.7(7) | 5907.27221 | 0.10396 | 0.01894 | 7A2 |
| 5915.86037(3) | 9304.19389(3) | 0.50(8) | 6.0(9) | 5915.9435 | -0.08312 | 0.008909 | 7A2 |
| 5917.11692(4) | 9305.45043(4) | 0.8(2) | 8(1) | 5917.18048 | -0.06356 | 0.01075 | 7A2 |
| 5921.47054(3) | 9309.80406(3) | 0.5(1) | 5(1) | 5921.56998 | -0.09944 | 0.006038 | 9A2 |
| 5925.15264(3) | 9313.48616(3) | 0.57(6) | 7.2(7) | 5925.21874 | -0.06610 | 0.007539 | 9A2 |
| 5929.18279(9) | 9317.51631(9) | 0.34(9) | 10(2) | 5929.18416 | -0.00137 | 0.003458 | 7A2 |
| 5932.07029(2) | 9320.40381(2) | 1.3(1) | 9.7(5) | 5932.17911 | -0.10882 | 0.01855 | 9A2 |
| 5934.79673(2) | 9323.13024(2) | 1.3(1) | 6.9(3) | 5935.40214 | -0.60541 | 0.02365 | 9A2 |
| 5936.64699(2) | 9324.98051(2) | 1.6(1) | 10.6(4) | 5936.90689 | -0.25990 | 0.01828 | 9A2 |
| 5938.42703(2) | 9326.76055(2) | 2.2(2) | 8.6(3) | 5938.75634 | -0.32931 | 0.03940 | 9A2 |
| 5951.75790(4) | 9340.09142(4) | 0.46(8) | 6(1) | 5951.75946 | -0.00156 | 0.008140 | 7A2 |
| 5969.19293(4) | 9357.52645(4) | 0.8(2) | 4(1) | 5969.52002 | -0.32709 | 0.01198 | 9A2 |
| 5971.22661(4) | 9359.56013(4) | 5(1) | 8(1) | 5971.55656 | -0.32995 | 0.07081 | 9A2 |
| 5979.27871(8) | 9367.61223(8) | 3.1(6) | 10(2) | 5979.45205 | -0.17333 | 0.07268 | 9A2 |

Table IX. Parameters of the lines detected in the Q(6, $F_2$)-pumped measurement, i.e. starting from the level with $J' = (6, F_1)$ and term value 219.9149048(8) + 3016.48912913(11) = 3236.4040339(8) cm$^{-1}$. Column 1 lists the measured center wavenumber. Column 2 shows the final state term value calculated as described in Section 4.1. Column 3 states the integrated absorption coefficient of the Lorentzian component of the probe transition. Column 4 shows the half width at half maximum of the Lorentzian component of the probe transition. Columns 5 and 6 show the predicted probe transition wavenumber from the effective Hamiltonian and its difference with respect to the observed wavenumber. Column 7 shows the Einstein A coefficient from the Hamiltonian. Column 8 shows the final state $J$ number and symmetry.

| 1 | 2 | 3 | 4 | 5 | 6 | 7 | 8 |
|---|---|---|---|---|---|---|---|
| Probe transition wavenumber [cm$^{-1}$] | Final state term value [cm$^{-1}$] | Probe transition integrated absorption coefficient [10$^{-9}$ cm$^{-2}$] | Probe transition width [MHz] | Hamiltonian transition wavenumber [cm$^{-1}$] | Obs. – pred. wavenumber [cm$^{-1}$] | Hamiltonian Einstein A coeff. [s$^{-1}$] | Final state $J$ number |
| 5841.01911(6) | 9077.42314(6) | 2.6(6) | 7(2) | 5841.10351 | -0.08441 | 0.07352 | 7F2 |
| 5866.61568(1) | 9103.01971(1) | 28(4) | 9.8(2) | 5866.74708 | -0.13140 | 0.8024 | 5F2 |
| 5883.8340(1) | 9120.2380(1) | 0.5(1) | 14(3) | 5883.75379 | 0.08022 | 0.007423 | 6F2 |
| 5884.40325(6) | 9120.80728(6) | 0.4(1) | 8(2) | 5884.39197 | 0.01128 | 0.007953 | 6F2 |
| 5886.85508(4) | 9123.25912(4) | 0.5(1) | 6(1) | 5886.86853 | -0.01345 | 0.01260 | 6F2 |
| 5927.02690(4) | 9163.43093(4) | 0.41(6) | 7(1) | 5927.25348 | -0.22658 | 0.01735 | 7F2 |
| 5927.26195(2) | 9163.66598(2) | 1.7(2) | 9.5(6) | 5927.57319 | -0.31125 | 0.04336 | 7F2 |
| 5927.79224(3) | 9164.19627(3) | 1.6(2) | 10.7(7) | 5928.13376 | -0.34153 | 0.01977 | 7F2 |
| 5932.43257(2) | 9168.83661(2) | 3.7(9) | 8.7(3) | 5933.05499 | -0.62242 | 0.1122 | 6F2 |
| 5951.46698(3) | 9187.87102(3) | 2.0(5) | 5.9(7) | 5951.8834 | -0.41641 | 0.05809 | 6F2 |
| 5956.54811(3) | 9192.95214(3) | 2.8(4) | 6.7(7) | 5956.72596 | -0.17785 | 0.07397 | 7F2 |

Table X. Term values of the final energy levels measured and assigned in this work. Column 1 gives the experimental final state term values, calculated as a weighted mean whenever the same level was reached in more than one measurement. Column 2 gives the rotational labels of the assignments while column 3 gives the counting numbers of the levels in the Hamiltonian, with the numbers from ExoMol in parentheses where they differ. Columns 4 and 5 display the predicted term values and vibrational assignments ($n_1$ $n_2$ $n_3$ $n_4$ sym) of the levels from the Hamiltonian. The corresponding term values and vibrational assignments ($n_1$ $n_2$ $l_2$ $n_3$ $n_4$ $l_4$ $m_4$ sym) from ExoMol are given in Columns 6 and 7. Column 8 indicates for which pump transitions the levels were reached. Note that the first level could not be assigned using the ExoMol line list.

| 1 | 2 | 3 | 4 | 5 | 6 | 7 | 8 |
|---|---|---|---|---|---|---|---|
| Measured term value [cm$^{-1}$] | Rotational assignment | Counting number | Predicted term value Hamiltonian [cm$^{-1}$] | Vibrational assignment Hamiltonian | Predicted term value ExoMol [cm$^{-1}$] | Vibrational assignment ExoMol | Observed via pump transitions |
| 9077.42314(6) | 7F2 | 935 | 9077.507444 | 0 1 2 1 F1 | // | // | Q(6, $F_2$) |
| 9103.01971(1) | 5F2 | 820 | 9103.151014 | 0 0 3 0 F1 | 9104.1233 | 0 0 0 3 3 3 0 0 0 F1 | Q(6,$F_2$) |
| 9120.2380(1) | 6F2 | 953 | 9120.157725 | 0 0 3 0 F2 | 9120.2391 | 0 0 0 3 1 1 0 0 0 F2 | Q(6,$F_2$) |
| 9120.80728(6) | 6F2 | 954 | 9120.795904 | 0 0 3 0 F2 | 9120.8087 | 0 0 0 3 1 1 0 0 0 F2 | Q(6,$F_2$) |
| 9123.25912(4) | 6F2 | 957 | 9123.272464 | 0 0 3 0 F2 | 9123.4625 | 0 0 0 3 1 1 0 0 0 F2 | Q(6,$F_2$) |
| 9158.60255(5) | 7A2 | 356 | 9158.893572 | 1 0 2 0 E | 9159.3327 | 0 0 0 3 3 2 0 0 0 E | P(7,$A_2$) |
| 9163.33053(3) | 7A2 | 357 | 9163.576 | 0 1 2 1 F2 | 9164.1874 | 0 1 1 2 2 2 1 1 1 F2 | P(7,$A_2$) |
| 9163.43093(4) | 7F2 | 1044 | 9163.657405 | 0 1 2 1 F1 | 9164.2234 | 0 1 1 2 2 2 1 1 1 F2 | Q(6,$F_2$) |
| 9163.66598(2) | 7F2 | 1045 | 9163.977125 | 0 1 2 1 F2 | 9164.6105 | 0 3 3 1 1 1 1 1 1 F1 | Q(6,$F_2$) |
| 9164.19627(3) | 7F2 | 1046 | 9164.537694 | 0 1 2 1 F2 | 9165.0838 | 0 1 1 2 2 2 1 1 1 F1 | Q(6,$F_2$) |
| 9164.79651(1) | 6A2 | 319 | 9164.990226 | 0 0 3 0 F1 | 9165.8792 | 0 0 0 3 3 3 0 0 0 F1 | P(7,$A_2$), Q(7,$A_2$) |
| 9165.78110(1) | 6A2 | 320 | 9165.927259 | 0 5 0 1 F2 | 9166.7122 | 0 0 0 3 3 3 0 0 0 F2 | P(7,$A_2$) |
| 9167.33214(5) | 5A2 | 291 | 9167.741555 | 0 2 2 0 F2 | 9167.5897 | 0 2 0 2 2 1 0 0 0 F2 | P(7,$A_2$) |
| 9168.83661(2) | 6F2 | 988 | 9169.458919 | 0 2 2 0 A1 | 9169.2579 | 0 2 0 2 0 0 0 0 0 A1 | Q(6,$F_2$) |
| 9183.18331(4) | 5A2 | 295 | 9183.562235 | 0 2 2 0 F1 | 9183.4082 | 0 2 2 2 2 1 0 0 0 F1 | P(7,$A_2$) |
| 9187.87102(3) | 6F2 | 995 | 9188.287329 | 0 2 2 0 E | 9188.548 | 0 2 2 2 0 0 0 0 0 E | Q(6,$F_2$) |
| 9191.32708(3) | 7A2 | 370 | 9191.465129 | 0 0 3 0 F2 | 9191.3284 | 0 0 0 3 1 1 0 0 0 F2 | P(7,$A_2$) |
| 9192.60911(1) | 7A2 | 371 | 9192.764637 | 0 0 3 0 F2 | 9192.934 | 0 0 0 3 1 1 0 0 0 F2 | P(7,$A_2$), Q(7,$A_2$) |
| 9192.95214(3) | 7F2 | 1079 | 9193.12989 | 0 0 3 0 F2 | 9193.2544 | 0 0 0 3 1 1 0 0 0 F2 | Q(6,$F_2$) |
| 9197.80501(1) | 5A2 | 296 | 9197.914549 | 0 0 3 0 F2 | 9197.8759 | 0 0 0 3 3 1 0 0 0 F2 | P(7,$A_2$) |
| 9198.67833(5) | 7A2 | 373 | 9198.955557 | 0 3 1 1 F1 | 9199.7741 | 0 0 0 3 1 1 0 0 0 F2 | P(7,$A_2$) |
| 9203.18941(6) | 7A2 | 374 | 9203.677742 | 0 3 1 1 E | 9204.0887 | 0 3 1 1 1 1 1 1 1 F1 | Q(7,$A_2$) |
| 9205.25178(1) | 5A2 | 297 | 9205.2524 | 0 0 3 0 F2 | 9205.2524 | 0 0 0 3 3 1 0 0 0 F2 | P(7,$A_2$) |
| 9206.25366(2) | 7A2 | 375 | 9206.617199 | 0 3 1 1 F2 | 9206.2532 | 0 0 0 3 1 1 0 0 0 F2 | Q(7,$A_2$) |
| 9209.21630(1) | 5A2 | 298 | 9209.582185 | 0 2 2 0 E | 9209.2167 | 0 2 0 2 2 2 0 0 0 E | P(7,$A_2$) |
| 9211.73907(7) | 7A2 | 378 | 9212.046126 | 0 3 1 1 F1 | 9212.4386 | 1 2 2 1 1 0 0 0 1 F1 | P(7,$A_2$) |
| 9214.28981(3) | 8A2 | 375 | 9214.386981 | 0 1 2 1 E | 9215.2552 | 0 1 1 2 2 1 1 1 1 E | Q(7,$A_2$), R(7,$A_2$) |
| 9216.25089(7) | 8A2 | 376 | 9216.327294 | 1 3 0 1 F2 | 9217.6904 | 2 2 2 0 0 0 0 0 0 E | R(7,$A_2$) |
| 9219.31581(3) | 7A2 | 380 | 9219.560301 | 0 3 1 1 F1 | 9220.2798 | 0 5 3 0 0 0 1 1 1 F1 | P(7,$A_2$) |
| 9219.48219(4) | 8A2 | 377 | 9219.71409 | 2 2 0 0 E | 9220.5063 | 0 1 1 2 2 1 1 1 1 F2 | R(7,$A_2$) |
| 9224.59833(2) | 7A2 | 381 | 9224.892724 | 0 3 1 1 F2 | 9225.9645 | 0 3 3 1 1 1 1 1 1 F1 | P(7,$A_2$) |
| 9224.92958(5) | 8A2 | 380 (379) | 9225.180148 | 1 0 2 0 E | 9225.119 | 0 0 0 3 3 2 0 0 0 E | R(7$\mathcal{A}_2$) |
| 9230.99155(2) | 6A2 | 330 | 9231.428925 | 0 2 2 0 F2 | 9231.2236 | 0 2 0 2 2 1 0 0 0 F2 | P(7$\mathcal{A}_2$) |
| 9232.13558(3) | 9A2 | 396 (394) | 9232.385911 | 1 3 0 1 F2 | 9233.6006 | 1 3 3 0 0 0 1 1 1 F2 | R(7$\mathcal{A}_2$) |
| 9232.496707(9) | 7A2 | 382 | 9232.670207 | 0 0 3 0 F1 | 9233.4705 | 0 0 0 3 3 3 0 0 0 F1 | P(7$\mathcal{A}_2$), R(7$\mathcal{A}_2$) |
| 9233.12918(3) | 8A2 | 383 | 9233.590813 | 1 0 2 0 E | 9233.3432 | 0 0 0 3 3 2 0 0 0 E | R(7$\mathcal{A}_2$) |
| 9233.18258(1) | 7A2 | 383 | 9233.400128 | 0 0 3 0 F1 | 9234.2549 | 0 0 0 3 3 3 0 0 0 F1 | P(7$\mathcal{A}_2$), Q(7$\mathcal{A}_2$), R(7$\mathcal{A}_2$) |
| 9241.535536(9) | 7A2 | 384 | 9241.863833 | 0 5 0 1 F1 | 9242.2168 | 0 2 0 2 0 0 0 0 0 A1 | P(7$\mathcal{A}_2$), Q(7$\mathcal{A}_2$) |
| 9242.58110(2) | 8A2 | 387 | 9242.987194 | 0 1 2 1 F1 | 9243.5322 | 0 1 1 2 2 2 1 1 1 F1 | Q(7$\mathcal{A}_2$) |
| 9245.58996(1) | 7A2 | 385 | 9245.852528 | 0 2 2 0 A1 | 9246.1039 | 0 2 0 2 0 0 0 0 0 A1 | Q(7$\mathcal{A}_2$) |
| 9246.72563(1) | 8A2 | 389 | 9246.957389 | 0 1 2 1 F1 | 9247.4409 | 0 1 1 2 2 2 1 1 1 F1 | Q(7$\mathcal{A}_2$), R(7$\mathcal{A}_2$) |
| 9249.96268(5) | 8A2 | 390 | 9249.718104 | 0 5 0 1 F2 | 9250.9888 | 1 2 0 1 1 1 0 0 0 F2 | Q(7$\mathcal{A}_2$) |
| 9250.45343(1) | 8A2 | 391 | 9250.72241 | 0 1 2 1 F1 | 9251.7711 | 0 1 1 2 2 2 1 1 1 F1 | Q(7$\mathcal{A}_2$), R(7$\mathcal{A}_2$) |
| 9254.40293(3) | 9A2 | 404 (403) | 9254.4383 | 0 1 2 1 F1 | 9255.3332 | 0 1 1 2 2 1 1 1 1 F1 | R(7$\mathcal{A}_2$) |
| 9255.03361(2) | 7A2 | 386 | 9255.041354 | 0 5 0 1 F1 | 9256.3571 | 0 5 5 0 0 0 1 1 1 F1 | P(7$\mathcal{A}_2$), Q(7$\mathcal{A}_2$) |
| 9255.93410(4) | 9A2 | 405 (404) | 9256.015996 | 2 0 1 0 F2 | 9256.8103 | 0 1 1 2 2 1 1 1 1 F2 | R(7$\mathcal{A}_2$) |
| 9256.26899(3) | 8A2 | 393 | 9256.654339 | 0 1 2 1 F2 | 9256.9652 | 1 2 0 1 1 1 0 0 0 F2 | Q(7$\mathcal{A}_2$) |
| 9262.149014(9) | 7A2 | 387 | 9262.319617 | 1 4 0 0 E | 9262.9045 | 0 2 2 2 0 0 0 0 0 E | P(7$\mathcal{A}_2$), Q(7$\mathcal{A}_2$), R(7$\mathcal{A}_2$) |
| 9262.61483(1) | 6A2 | 335 | 9262.6157 | 0 0 3 0 F2 | 9262.6156 | 0 0 0 3 3 1 0 0 0 F2 | P(7$\mathcal{A}_2$), Q(7$\mathcal{A}_2$) |
| 9264.709372(9) | 7A2 | 388 | 9265.045873 | 0 2 2 0 E | 9265.4895 | 0 2 2 2 0 0 0 0 0 E | P(7$\mathcal{A}_2$), Q(7$\mathcal{A}_2$), R(7$\mathcal{A}_2$) |
| 9265.90687(3) | 9A2 | 410 (409) | 9266.238041 | 0 3 1 1 F2 | 9266.9864 | 0 1 1 2 2 1 1 1 1 F2 | R(7$\mathcal{A}_2$) |
| 9268.85049(4) | 8A2 | 398 (399) | 9268.453901 | 0 5 0 1 F2 | 9270.391 | 0 5 5 0 0 0 1 1 1 F2 | Q(7$\mathcal{A}_2$), R(7$\mathcal{A}_2$) |
| 9269.93886(3) | 8A2 | 399 (398) | 9270.259374 | 1 2 1 0 F2 | 9269.5618 | 0 5 5 0 0 0 1 1 1 F2 | R(7$\mathcal{A}_2$) |
| 9270.34839(7) | 9A2 | 412 (411) | 9270.607268 | 0 3 1 1 F2 | 9271.3611 | 0 1 1 2 2 1 1 1 1 F2 | R(7$\mathcal{A}_2$) |
| 9272.383867(8) | 7A2 | 389 | 9272.270247 | 1 4 0 0 A1 | 9273.3088 | 1 0 0 2 2 0 0 0 0 A1 | P(7$\mathcal{A}_2$), Q(7$\mathcal{A}_2$), R(7$\mathcal{A}_2$) |
| 9272.74257(1) | 8A2 | 400 | 9272.621238 | 0 5 0 1 F2 | 9273.2997 | 0 0 0 3 1 1 0 0 0 F2 | Q(7$\mathcal{A}_2$), R(7$\mathcal{A}_2$) |
| 9273.59978(2) | 6A2 | 336 | 9274.084639 | 0 2 2 0 E | 9273.6735 | 0 2 0 2 2 0 0 0 0 E | P(7$\mathcal{A}_2$) |
| 9274.64132(2) | 8A2 | 401 | 9274.656812 | 0 0 3 0 F2 | 9275.1026 | 0 0 0 3 1 1 0 0 0 F2 | Q(7$\mathcal{A}_2$) |
| 9278.666752(9) | 7A2 | 390 | 9278.768498 | 0 5 0 1 F1 | 9279.5538 | 0 5 1 0 0 0 1 1 1 F2 | P(7$\mathcal{A}_2$), Q(7$\mathcal{A}_2$), R(7$\mathcal{A}_2$) |
| 9280.39994(1) | 8A2 | 403 | 9280.716122 | 1 2 1 0 F2 | 9280.7387 | 0 0 0 3 1 1 0 0 0 F2 | Q(7$\mathcal{A}_2$), R(7$\mathcal{A}_2$) |
| 9282.65976(4) | 6A2 | 338 | 9283.065221 | 0 2 2 0 A2 | 9282.9602 | 0 2 2 2 2 2 0 0 0 A2 | P(7$\mathcal{A}_2$) |
| 9284.29985(3) | 8A2 | 405 | 9284.649497 | 0 3 1 1 F1 | 9285.3557 | 0 3 3 1 1 1 1 1 1 F1 | R(7$\mathcal{A}_2$) |
| 9286.312823(8) | 7A2 | 391 | 9286.377142 | 0 2 2 0 E | 9286.9015 | 0 2 2 2 0 0 0 0 0 E | P(7$\mathcal{A}_2$), Q(7$\mathcal{A}_2$), R(7$\mathcal{A}_2$) |
| 9287.44658(5) | 8A2 | 406 | 9287.632352 | 0 3 1 1 E | 9288.4015 | 0 3 1 1 1 1 1 1 1 E | Q(7$\mathcal{A}_2$) |
| 9295.70967(1) | 7A2 | 392 | 9295.605456 | 0 5 0 1 F2 | 9296.5403 | 0 2 0 2 0 0 0 0 0 A1 | P(7$\mathcal{A}_2$), Q(7$\mathcal{A}_2$), R(7$\mathcal{A}_2$) |
| 9302.00947(6) | 7A2 | 394 | 9302.3361 | 0 2 2 0 F2 | 9302.389 | 0 2 0 2 2 1 0 0 0 F2 | P(7$\mathcal{A}_2$) |
| 9304.19393(1) | 7A2 | 395 | 9304.276737 | 0 5 0 1 F2 | 9304.805 | 0 2 2 2 0 0 0 0 0 E | P(7$\mathcal{A}_2$), R(7$\mathcal{A}_2$) |
| 9305.45054(1) | 7A2 | 396 | 9305.513723 | 1 4 0 0 E | 9306.3874 | 0 2 0 2 0 0 0 0 0 A1 | P(7$\mathcal{A}_2$), Q(7$\mathcal{A}_2$), R(7$\mathcal{A}_2$) |
| 9309.80406(3) | 9A2 | 435 (434) | 9309.903217 | 0 1 2 1 F2 | 9311.2506 | 0 1 1 2 2 1 1 1 1 A2 | R(7$\mathcal{A}_2$) |
| 9313.48616(3) | 9A2 | 436 (435) | 9313.551979 | 0 1 2 1 F2 | 9315.4507 | 0 1 1 2 2 1 1 1 1 F1 | R(7$\mathcal{A}_2$) |
| 9314.77055(5) | 7A2 | 398 | 9315.074687 | 0 2 2 0 F1 | 9314.9155 | 0 2 2 2 2 1 0 0 0 F1 | Q(7$\mathcal{A}_2$) |
| 9314.91940(1) | 8A2 | 415 | 9315.195488 | 0 0 3 0 F1 | 9315.9929 | 0 0 0 3 3 3 0 0 0 F1 | Q(7$\mathcal{A}_2$) |
| 9317.51603(2) | 7A2 | 399 | 9317.5174 | 0 2 2 0 F2 | 9317.5173 | 0 2 2 2 2 1 0 0 0 F2 | P(7,$\mathcal{A}_2$), Q(7,$\mathcal{A}_2$), R(7,$\mathcal{A}_2$) |
| 9320.40381(2) | 9A2 | 439 (438) | 9320.512354 | 0 1 2 1 F2 | 9321.7194 | 0 1 1 2 2 1 1 1 1 F2 | R(7,$\mathcal{A}_2$) |
| 9323.13024(2) | 9A2 | 440 (439) | 9323.735376 | 1 0 2 0 E | 9323.9038 | 0 0 0 3 3 2 0 0 0 E | R(7,$\mathcal{A}_2$) |
| 9324.98051(2) | 9A2 | 441 (440) | 9325.240132 | 0 3 1 1 F2 | 9326.0352 | 0 3 3 1 1 1 1 1 1 F2 | R(7,$\mathcal{A}_2$) |
| 9326.76055(2) | 9A2 | 442 (441) | 9327.089581 | 1 0 2 0 E | 9327.5682 | 0 0 0 3 3 2 0 0 0 E | R(7,$\mathcal{A}_2$) |
| 9326.89457(3) | 7A2 | 402 | 9326.939424 | 1 4 0 0 E | 9327.3179 | 0 2 2 2 0 0 0 0 0 E | Q(7,$\mathcal{A}_2$) |
| 9332.90967(1) | 7A2 | 403 | 9333.05949 | 0 0 3 0 F2 | 9333.018 | 0 0 0 3 3 1 0 0 0 F2 | P(7,$\mathcal{A}_2$), Q(7,$\mathcal{A}_2$) |
| 9340.09138(1) | 7A2 | 404 | 9340.0927 | 0 0 3 0 F2 | 9340.0926 | 0 0 0 3 3 1 0 0 0 F2 | P(7,$\mathcal{A}_2$), Q(7,$\mathcal{A}_2$), R(7,$\mathcal{A}_2$) |
| 9348.13775(6) | 7A2 | 405 | 9348.1388 | 0 2 2 0 E | 9348.1376 | 0 2 0 2 2 2 0 0 0 E | Q(7,$\mathcal{A}_2$) |
| 9357.52645(4) | 9A2 | 460 (459) | 9357.853258 | 0 0 3 0 F2 | 9358.4089 | 0 1 1 2 2 2 1 1 1 F2 | R(7,$\mathcal{A}_2$) |
| 9359.56013(4) | 9A2 | 462 (461) | 9359.8898 | 1 2 1 0 F2 | 9360.2176 | 0 5 5 0 0 0 1 1 1 F2 | R(7,$\mathcal{A}_2$) |
| 9367.61223(8) | 9A2 | 464 (463) | 9367.785288 | 1 0 2 0 F2 | 9367.8317 | 0 0 0 3 1 1 0 0 0 F2 | R(7,$\mathcal{A}_2$) |

Table XI. The observed V-type transitions from the vibrational ground state with Hamiltonian assignments. Column 1 lists the observed center wavenumber. Columns 2 and 3 give the center wavenumbers and line intensities predicted by the effective Hamiltonian. Columns 4 to 6 give rotational assignments, counting numbers and vibrational assignments ($n_1$ $n_2$ $n_3$ $n_4$ sym) for the upper state. Column 7 gives the lower rotational assignments.

| 1 | 2 | 3 | 4 | 5 | 6 | 7 |
|---|---|---|---|---|---|---|
| Transition wavenumber [cm$^{-1}$] | Hamiltonian transition wavenumber [cm$^{-1}$] | Hamiltonian line intensity [cm$^{-1}$/(mol cm$^{-2}$)] | Upper state $J$-number | Upper state counting number | Vibrational assignment | Lower state $J$-number |
| 5842.26729(7) | 5842.265118 | $1.323 \cdot 10^{-23}$ | 7A1 | 56 | 0 1 1 1 0F1 | 7A2 |
| 5844.25859(5) | 5844.260815 | $1.293 \cdot 10^{-23}$ | 7A1 | 57 | 0 3 0 1 0F1 | 7A2 |
| 5859.94619(5) | 5859.946726 | $3.06 \cdot 10^{-23}$ | 7A1 | 61 | 1 0 1 0 0F2 | 7A2 |
| 5862.38978(6) | 5862.389971 | $5.066 \cdot 10^{-24}$ | 6F1 | 164 | 0 1 1 1 0F1 | 6F2 |
| 5864.28352(4) | 5864.282265 | $8.163 \cdot 10^{-23}$ | 8A1 | 63 | 0 1 1 1 0F1 | 7A2 |
| 5864.91925(6) | 5864.919605 | $3.228 \cdot 10^{-23}$ | 7A1 | 62 | 0 1 1 1 0F1 | 7A2 |
| 5869.81539(6) | 5869.81456 | $1.83 \cdot 10^{-23}$ | 7F1 | 166 | 0 1 1 1 0F1 | 6F2 |
| 5870.82610(6) | 5870.82507 | $1.136 \cdot 10^{-23}$ | 6F1 | 171 | 0 1 1 1 0F2 | 6F2 |
| 5871.27120(3) | 5871.271878 | $5.705 \cdot 10^{-23}$ | 7A1 | 63 | 0 1 1 1 0A2 | 7A2 |
| 5875.22392(4) | 5875.22075 | $8.707 \cdot 10^{-23}$ | 8A1 | 65 | 0 1 1 1 0F1 | 7A2 |
| 5880.35099(6) | 5880.352688 | $1.217 \cdot 10^{-23}$ | 8A1 | 66 | 0 1 1 1 0F2 | 7A2 |
| 5885.04837(6) | 5885.043824 | $1.517 \cdot 10^{-23}$ | 8A1 | 67 | 2 0 0 0 0A1 | 7A2 |
| 5891.50059(5) | 5891.49858 | $1.053 \cdot 10^{-22}$ | 7F1 | 171 | 0 1 1 1 0F1 | 6F2 |
| 5910.12483(6) | 5910.125312 | $1.295 \cdot 10^{-22}$ | 8A1 | 70 | 0 1 1 1 0F1 | 7A2 |
| 5930.59480(5) | 5930.593091 | $5.471 \cdot 10^{-24}$ | 8A1 | 74 | 1 0 1 0 0F2 | 7A2 |
| 5960.24637(6) | 5960.245259 | $9.829 \cdot 10^{-24}$ | 8A1 | 79 | 0 3 0 1 0F1 | 7A2 |
| 5961.71838(7) | 5961.714536 | $2.861 \cdot 10^{-23}$ | 8A1 | 80 | 0 1 1 1 0F2 | 7A2 |
| 5968.85679(5) | 5968.856148 | $1.702 \cdot 10^{-23}$ | 6A1 | 72 | 0 0 2 0 0E | 7A2 |
| 5969.62118(5) | 5969.621881 | $4.634 \cdot 10^{-24}$ | 7A1 | 71 | 1 2 0 0 0E | 7A2 |
| 5985.42358(6) | 5985.422914 | $2.995 \cdot 10^{-24}$ | 6A1 | 73 | 0 2 1 0 0F2 | 7A2 |
| 5988.81868(6) | 5988.819078 | $1.119 \cdot 10^{-23}$ | 6A1 | 74 | 0 2 1 0 0F2 | 7A2 |
| 5993.38190(4) | 5993.381143 | $1.597 \cdot 10^{-23}$ | 6A1 | 76 | 0 2 1 0 0F2 | 7A2 |
| 6006.06681(6) | 6006.066032 | $6.742 \cdot 10^{-24}$ | 6A1 | 77 | 0 2 1 0 0F2 | 7A2 |
| 6039.65774(7) | 6039.658145 | $3.577 \cdot 10^{-23}$ | 7A1 | 74 | 0 0 2 0 0E | 7A2 |
| 6053.82527(7) | 6053.827205 | $1.864 \cdot 10^{-24}$ | 8A1 | 89 | 1 2 0 0 0E | 7A2 |
| 6061.91307(4) | 6061.915178 | $2.532 \cdot 10^{-23}$ | 7A1 | 76 | 0 2 1 0 0F2 | 7A2 |
| 6070.02341(3) | 6070.023553 | $1.55 \cdot 10^{-23}$ | 7A1 | 78 | 0 2 1 0 0F1 | 7A2 |